\documentclass{article}

\usepackage[preprint]{neurips_2025}

\usepackage{graphicx}
\usepackage[utf8]{inputenc} %
\usepackage[T1]{fontenc}    %
\usepackage{hyperref}       %
\usepackage{url}            %
\usepackage{booktabs}       %
\usepackage{amsfonts}       %
\usepackage{amssymb}
\usepackage{nicefrac}       %
\usepackage{microtype}      %
\usepackage{xcolor}         %
\usepackage{lipsum}
\usepackage{xspace}
\usepackage{longtable}
\usepackage{lscape}
\usepackage{subcaption}
\usepackage{amsmath}
\usepackage{multirow}
\usepackage{pifont}
\usepackage[hang,flushmargin]{footmisc}
\let\oldding\ding%
\renewcommand{\ding}[2][1]{\scalebox{#1}{\oldding{#2}}}

\usepackage{siunitx} %
\usepackage{caption} %

\newcommand{\name}{\textsc{DroneAudioset}\xspace}

\newcommand{\dronelarge}{$\mathbb{D}_{large}$\xspace}
\newcommand{\dronesmall}{$\mathbb{D}_{small}$\xspace}
\newcommand{\micup}{$\mathbb{M}_{up}$\xspace}
\newcommand{\micdown}{$\mathbb{M}_{down}$\xspace}
\newcommand{\miccenter}{$\mathbb{M}_{center}$\xspace}
\newcommand{\recordingduration}{23.5\xspace}

\newcommand{\sectioncolor}{violet}
\newcommand{\cmark}{\ding[1.2]{51}}%
\newcommand{\xmark}{\color{red}{\ding[1.2]{55}}}%

\title{\name: An Audio Dataset for Drone-based Search and Rescue}

\author{
  \normalfont Chitralekha Gupta\textsuperscript{*} \and Soundarya Ramesh\textsuperscript{*} \and Praveen Sasikumar \and Kian Peen Yeo \and Suranga Nanayakkara\\
  \textsuperscript{*}Equal contribution \\
  School of Computing, National University of Singapore \\
  \texttt{\{chitralekha, soundarya, praveen, kpyeo, suranga\}@ahlab.org}
}

\begin{document}

\maketitle
\begin{abstract}
Unmanned Aerial Vehicles (UAVs) or \textit{drones}, are increasingly used in search and rescue missions to detect human presence. Existing systems primarily leverage vision-based methods which are prone to fail under low-visibility or occlusion. Drone-based audio perception offers promise but suffers from extreme ego-noise that masks sounds indicating human presence.  
Existing datasets are either limited in diversity or synthetic, lacking real acoustic interactions, and there are no standardized setups for drone audition. 
To this end, we present \name\footnote{The dataset is publicly available at \url{https://huggingface.co/datasets/ahlab-drone-project/DroneAudioSet/} under the MIT license.\\Code is available at \url{https://github.com/augmented-human-lab/DroneAudioSet-code.git}.\\Webpage: \url{https://apps.ahlab.org/DroneAudioSet-code/}.
}, a comprehensive drone audition dataset featuring \recordingduration hours of annotated recordings, covering a wide range of signal-to-noise ratios (SNRs) from -57.2 dB to -2.5 dB, across various drone types, throttles, microphone configurations as well as environments. The dataset enables development and systematic evaluation of noise suppression and classification methods for human-presence detection under challenging conditions, while also informing practical design considerations for drone audition systems, such as microphone placement trade-offs, and development of drone noise-aware audio processing. This dataset is an important step towards enabling design and deployment of drone-audition systems.

\end{abstract}

\vspace{-0.2cm}
\section{Introduction}
\label{sec:introduction}
\vspace{-0.2cm}
Unmanned Aerial Vehicles (UAVs), commonly known as \textit{drones}, have become invaluable tools for search and rescue (SAR), environmental monitoring, and surveillance. Specifically, for SAR operations in \textit{indoor environments}, such as dilapidated buildings due to earthquake and fire accidents, drones are indispensable in identifying human presence to inform rescue operators. Traditionally, these missions rely on visual data which face challenges in low-visibility conditions (e.g., smoke, fog, cluttered environments) due to occlusions, reflections, as well as bandwidth constraints for video transmission \cite{martinez2020review}. Auditory scene analysis offers a complementary modality that penetrates visual obstructions, enabling detection of human presence through sounds such as human vocal sounds (e.g.~speech, scream and cries), as well as non-verbal cues such as banging and footsteps. However, the potential of drone-mounted microphones remains underexplored due to two key challenges. First, drones generate intense noise from motors and propellers, known as ego-noise, combined with wind noise \cite{martinez2020review,deleforge2020drone}\footnote{Ego-noise of drones often exceeds 80~dBA at a 1-meter distance. Refer to Table \ref{tbl:drone-throttle} in the Appendix.} %, included as part of supplementary material.}. 
This noise overlaps and masks human vocal frequencies, %resulting in 
significantly degrading the signal-to-noise ratio (SNR) to below -10~dB \cite{deleforge2020drone}, making human presence detection difficult. 
Second, this ego-noise is spatially uneven due to turbulence from the propellers, causing some microphone placements, when mounted on or around the drone, to be more exposed to wind noise than others. Yet, there is limited understanding of how different microphone positions, drone throttles, and source types affect recording quality \cite{martinez2020review,deleforge2020drone}. These challenges highlight the need for a comprehensive drone audio dataset to systematically study ego-noise characteristics and evaluate microphone placement strategies in diverse acoustic conditions for effective human presence detection.

A few existing datasets capture human sounds using drone-mounted microphones. Among them, the DRone EGO-Noise (DREGON) dataset \cite{strauss2018dregon} provides indoor recordings for speech localization tasks. However, its scope is limited in terms of amount of recording data, as well as diversity in SNR. %
Synthetic datasets \cite{morito2016partially,morito2016reduction} that combine a pre-recorded drone noise at a specific setting with clean source audio are unable to reflect real-world acoustic interactions and variability, particularly the dynamic modulation of rotor harmonics with throttle changes and the effects of wind turbulence.%

To address these limitations, we introduce \name, a systematically collected drone audio dataset comprising \recordingduration hours of high-quality recordings. The data was collected using a controlled experimental setup in which the drone was securely mounted on a fixed frame, ensuring consistent and repeatable conditions. This methodology allowed us to capture a diverse range of annotated audio samples across multiple critical parameters, including different microphones, varying drone sizes and throttle settings, diverse acoustic environments, and a wide array of sound sources relevant to search and rescue operations. 
Overall, our dataset captures source sounds in combination with drone noise, for a wide range of signal-to-noise ratios (SNRs) from -57.2 dB to -2.5 dB\footnote{0 dB SNR indicates that the noise and signal have equal loudness, and  every 3~dB decrease in SNR indicates doubling of the noise power.}.
Availability of such a dataset that captures the challenging acoustic scenarios in drone audition would boost development and evaluation of robust noise suppression and classification models for drone-audition based search and rescue. Moreover, this dataset could support system design for drone-audition, enabling empirical evaluation of hardware and operational design decisions, such as microphone placement, drone size, or optimal throttle levels for specific detection ranges.

Specifically, we make three contributions:\\
\textbullet~We present the first publicly available dataset of drone audio recordings, comprising \recordingduration hours of data captured under systematically controlled conditions. Our collection methodology varies key parameters including throttle levels, microphone configurations, source characteristics (type, volume, distance), and room acoustics to enable robust model development.\\
\textbullet~We provide a systematic evaluation of state-of-the-art noise suppression and audio classification models for human presence detection in drone environments, revealing fundamental limitations of current approaches under extreme low SNR conditions of in-flight recordings.\\
\textbullet~Through empirical analysis, we derive actionable recommendations for drone-audition systems, including optimal microphone placement strategies and operational parameters that balance acoustic performance with flight constraints.

Our dataset makes a significant contribution to the research community focused on drone audition and audio AI, by providing a dataset to understand the performance of auditory scene analysis methods under extreme noise conditions.

\vspace{-0.2cm}
\section{Related Work}
\label{sec:bg-and-related-work}
\vspace{-0.2cm}

\subsection{Existing Drone Audition Datasets}
\label{sec:related-work-drone-datasets}
\vspace{-0.3cm}
Researchers have proposed several drone audition datasets that focus on detecting and fingerprinting intruder drones~\cite{ruiz2018aira, kolamunna2021droneprint}, ground surface classification for disaster assessment~\cite{yano2024ground}, as well as identifying human sounds such as speech~\cite{strauss2018dregon, morito2016partially, wang2019audio}. With regards to capturing human sounds, DREGON \cite{strauss2018dregon} dataset offers in-flight recordings with human speech, white noise, and chirps, for supporting sound source localization. Their dataset consists of recordings acquired with a microphone array attached under the drone in an indoor environment, with a loudspeaker transmitting speech from different azimuth angles. 
Another approach, AVQ \cite{wang2019audio} similarly combines audio-visual data to track moving and static human speakers. However, both these datasets focus on the problem of source localization, while they remain limited in scenario diversity and sound type coverage that are crucial for realistic search-and-rescue audio modeling. Another line of work by Morito \textit{et al}.~\cite{morito2016partially} aims to address source sound identification and noise suppression. However, the dataset used relied on synthetic data, i.e.~the drone noise and source sounds were separately recorded and later convolved, rather than capturing both together in real-world scenarios. Additionally, the sound mixtures were combined at 0~dB SNR, which is far from realistic conditions where human-relevant sounds often lie below -10 dB under drone noise \cite{deleforge2020drone,strauss2018dregon,wang2019audio}. %
The dataset is also not publicly available, further restricting its usability for broader research efforts. 
To bridge this gap, we develop the \name dataset to provide extensive, real-world recordings across various environments, distances, and sound types, supporting robust human presence detection and audio scene analysis under drone noise. We provide a detailed comparison of the above closely related datasets in Table \ref{tbl:datasets}.

\begin{table}[]
    \centering
    \renewcommand{\arraystretch}{1}
    \resizebox{0.9\textwidth}{!}{
\begin{tabular}{|c|c|c|cc|c|c|c|c|}
\hline
Dataset &
  \begin{tabular}[c]{@{}c@{}}Dataset\\ Type\end{tabular} &
  \begin{tabular}[c]{@{}c@{}}No. of\\ Drones\end{tabular} &
  \multicolumn{2}{c|}{Source Types} &
  \begin{tabular}[c]{@{}c@{}}No. of\\ Mics\end{tabular} &
  SNR Diversity &
  \begin{tabular}[c]{@{}c@{}}Recording\\ Duration\end{tabular} &
  \begin{tabular}[c]{@{}c@{}}Publicly\\ Available\end{tabular} \\ \hline \hline
Morito \textit{et al.}~\cite{morito2016partially} &
  Synthetic &
  1 &
  \multicolumn{1}{c}{\begin{tabular}[c]{@{}c@{}}HV\\ HNV\\ NH\end{tabular}} &
  \begin{tabular}[c]{@{}c@{}}\cmark\\ \cmark\\ \cmark\end{tabular} &
  8 &
  0 dB &
  (not available) &
  \xmark \\ \hline
DREGON \cite{strauss2018dregon} &
  Real-world &
  1 &
  \multicolumn{1}{c}{\begin{tabular}[c]{@{}c@{}}HV\\ HNV\\ NH\end{tabular}} &
  \begin{tabular}[c]{@{}c@{}}\cmark\\ \xmark\\ \cmark\end{tabular} &
  8 &
  -16 to -10 dB &
  6.12 hours &
  \cmark \\ \hline
AVQ \cite{wang2019audio} &
  Real-world &
  1 &
  \multicolumn{1}{c}{\begin{tabular}[c]{@{}c@{}}HV\\ HNV\\ NH\end{tabular}} &
  \begin{tabular}[c]{@{}c@{}}\cmark\\ \xmark\\ \xmark\end{tabular} &
  8 &
  -32 to -15 dB & %-15 to -32 dB &
  0.84 hours &
  \xmark \\ \hline
\begin{tabular}[c]{@{}c@{}}\textbf{\name}\\\textbf{(Ours)}\end{tabular} &
  \textbf{Real-world} &
  \textbf{2} &
  \multicolumn{1}{c}{\begin{tabular}[c]{@{}c@{}}\textbf{HV}\\\textbf{HNV}\\\textbf{NH}\end{tabular}} &
  \begin{tabular}[c]{@{}c@{}}\cmark\\ \cmark\\ \cmark\end{tabular} &
  \textbf{17} &
  \textbf{-57.2 dB to -2.5 dB} &
  \textbf{\recordingduration hours} &
  \cmark \\ \hline
\end{tabular}}
\vspace{0.1cm}
\caption{Comparison of existing drone audition datasets with ours on the basis of dataset type, number of tested drones, sound source types, microphones, recording diversity and duration. Here, HV, HNV and NH refer to human vocal, human non-vocal and non-human (ambient) sounds, respectively.}
\label{tbl:datasets}
\vspace{-0.6cm}
\end{table}

\vspace{-0.2cm}
\subsection{Audio Event Detection Datasets in Non-Drone Contexts}
\label{sec:related-work-audio-event-datasets}
\vspace{-0.3cm}
There are several well-known datasets such as STARSS22 \cite{politis2022starss22}, FSD50k \cite{fonseca2022FSD50K}, and DESED \cite{serizel2020sound,turpault2019sound}, that are commonly used for benchmarking performance on tasks such as sound event detection in domestic environments, and localization in spatial auditory scenes in the DCASE community\footnote{\url{https://dcase.community/}}. These datasets contain common sound events from the AudioSet ontology \cite{gemmeke2017audioset}, including human sounds such as laughter, crying, speech, and non-human sounds such as telephone rings, mixed with ambient noise. While valuable for audio event detection research, these scenarios assume moderate SNRs, i.e., 14-26 dB on average \cite{fonseca2022FSD50K}), as well as relatively stable noise characteristics that differ significantly from drone environments. For speech processing, the Deep Noise Suppression (DNS) Challenges\footnote{\url{https://aka.ms/dns-challenge}} \cite{reddy2020interspeech} have established standardized evaluation using synthetic and real-world noisy speech datasets with SNRs ranging from 0 dB to 40 dB \cite{dubey2023icassp,reddy2020interspeech}. While these datasets cover challenging conditions such as babble noise and reverberation, they do not capture the extreme low-SNR scenarios (< -10 dB) characteristic of drone recordings where target speech competes with broadband rotor noise \cite{deleforge2020drone}.

\vspace{-0.2cm}
\section{\name Dataset}
\label{sec:dataset}
\vspace{-0.2cm}
\begin{figure}
    \centering
    \includegraphics[width=1\linewidth]{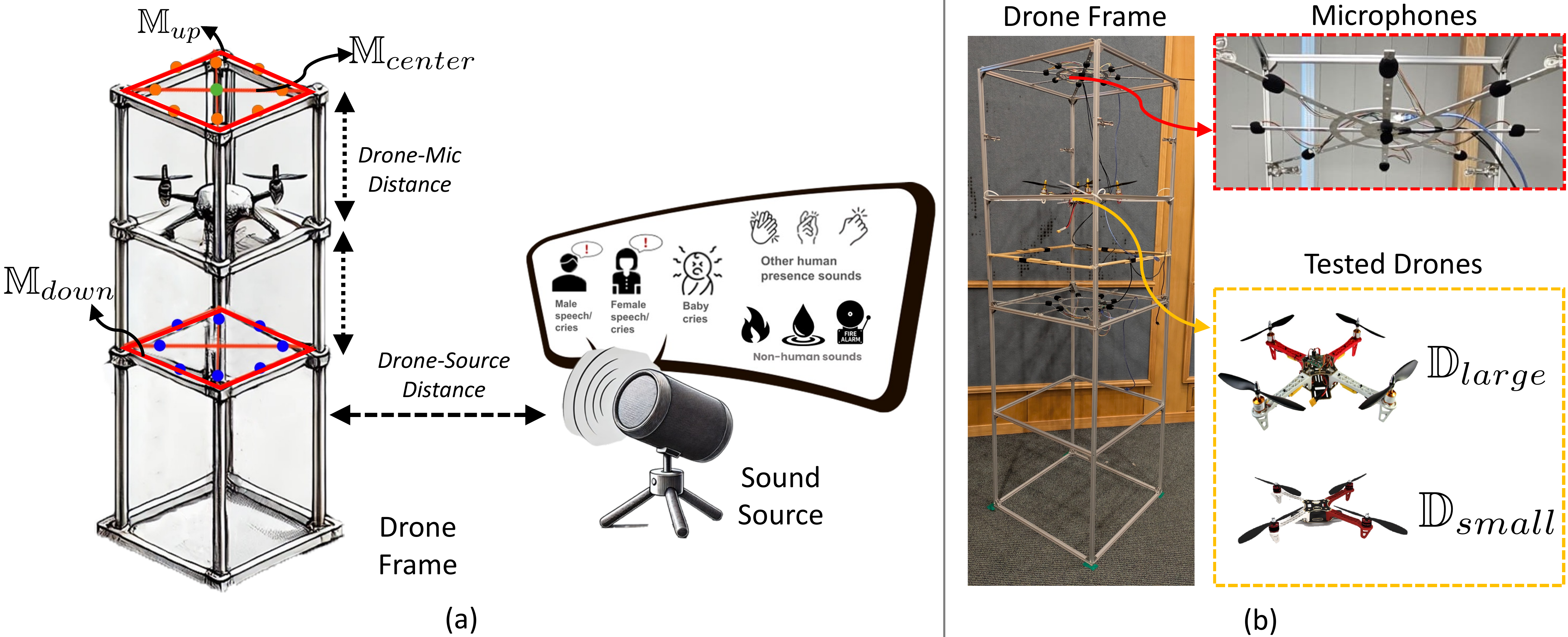}
    \vspace{-0.5cm}
    \caption{Figure (a) illustrates our experimental setup with the drone attached to a fixed aluminum frame, with two microphone arrays, \micup and \micdown, and a single microphone, \miccenter. The source sounds (i.e., human vocal sounds, human presence sounds and ambient sounds) are transmitted through a speaker. (b) the actual setup -- the drone frame, microphone array, and the drones used.}
    \label{fig:setup}
    \vspace{-0.5cm}
\end{figure}

We introduce a novel search-and-rescue drone audition dataset designed to identify human presence sounds by emulating realistic drone recording settings. To ensure diversity, our dataset incorporates variations in drone attributes (types and throttle speeds), microphone configurations (array setups and placements), and human sounds (types, locations and loudness). %

\vspace{-0.3cm}
\subsection{Data Collection Overview}
\label{ref:sec-dataset-setup}
\vspace{-0.3cm}
Figure~\ref{fig:setup}(a) depicts our experimental setup with an aluminum frame structure housing the drone. Similar to an ideal search-and-rescue setting where the drone hovers (while remaining static in air), we affix the drone to the frame to emulate the scenario, while also increasing flexibility for experimenting different recording configurations. We envision that in a real scenario, the microphones will be connected to drones, either directly affixed or through a solid structure, both of which will involve mechanical coupling between the drone and the microphones, resulting in structural vibrations being captured by the microphones. Hence, we created a setup with a common frame between the two. In addition, throughout the recording sessions, we ensured that the drone remained securely and evenly attached to the frame to prevent any anomalous vibrations. Figure~\ref{fig:setup}(b) shows the actual setup image of the frame, as well as the microphones and drones used. While most of our dataset includes audio data collected while simultaneously operating the drone and transmitting source sounds, we also collect \textit{drone-only} and \textit{source-only} audio data, for signal-to-noise ratio (SNR) computations (see Section~\ref{sec:dataset-snr}). In Table~\ref{tbl:data-collection-summary}, we enumerate all data recording configurations such as different drones, throttle levels, microphones, drone-microphone distances, sound sources, drone-source distances, source loudness levels, as well as environments. To the best of our knowledge, this is the largest search-and-rescue-focused drone audition dataset, with \recordingduration hours of high-quality audio recordings.

\vspace{-0.2cm}
\subsection{Drone Attributes}
\label{sec:dataset-drones}
\vspace{-0.3cm}
We experiment with quadcopters, consisting of four sets of motors and propellers, which are suited for indoor search-and-rescue settings given their quick maneuverability and vertical takeoff and landing capabilities~\cite{fixed-wing-vs-quadcopter-1, fixed-wing-vs-quadcopter-2}. 
In particular, we use DJI F450~\cite{dji-f450} and DJI F330~\cite{dji-f330} drones, with 45~cm and 33~cm \textit{wheelbase} (i.e., diagonal motor-to-motor distance), henceforth referred to as \dronelarge and \dronesmall, respectively. 
We chose drones of different sizes due to their varied ego-noise profiles, providing diversity to our audition dataset (see Table~\ref{tbl:drone-specs} in Appendix for specification comparison of the two drones). %
We collect data in two throttle speeds, `low' and `high' (sound pressure levels of the two speeds is provided in Appendix Table \ref{tbl:drone-throttle}), while affixing the drone firmly to the aluminum frame, to emulate the hover mode. The drone is affixed to the frame at a height of 1.5~m from the ground. Please refer to Appendix~\ref{app:dataset-additional-info} for detailed drone noise profiles.

\vspace{-0.2cm}
\subsection{Microphone Attributes}
\label{sec:dataset-microphones}
\vspace{-0.3cm}
To systematically study drone acoustics in the absence of standard microphone placement guidelines, we deploy 17 microphones in a flexible experimental configuration. This includes two 8-channel circular arrays (\micup and \micdown, above and below the drone respectively) using ICS-43434 microphones\footnote{\url{https://invensense.tdk.com/products/ics-43434/}} and a central standalone mic\footnote{\url{https://soundskrit.ca/}} above the drone (\miccenter), all mounted on a shared frame with the drone (Fig.~\ref{fig:setup}). The frequency response of each microphone of the 8-channel microphone arrays is 60~Hz to 20~kHz, and that of the central standalone microphone is 80~Hz to 10~kHz. %
We configure our arrays at 25~cm and 50~cm from the drone, to enable array configuration experimentation, while maintaining coupling with the drone, similar to a real-world setup. All microphones were covered with a microphone foam to reduce noise, especially wind noise.

The two 8-microphone arrays provide spatial acoustic diversity, with microphone channels strategically placed under, above, and between propellers. The microphone array diameters scale with drone size (50 cm for \dronelarge, 30 cm for \dronesmall), matching their wheelbases (45 cm/33 cm) to preserve center of gravity. The complete assembly of a microphone array (together with its mounting plate) weighs 270 grams, which is under each drone's payload capacity (Table~\ref{tbl:drone-specs}), ensuring flight viability.

\begin{table}[]
\centering
\renewcommand{\arraystretch}{1}
    \resizebox{0.9\textwidth}{!}{
\begin{tabular}{ccccccccl|}
\hline
\multicolumn{1}{|c|}{Drone} &
  \multicolumn{1}{c|}{\begin{tabular}[c]{@{}c@{}}Throttle\\ Mode\end{tabular}} &
  \multicolumn{1}{c|}{Microphones} &
  \multicolumn{1}{c|}{\begin{tabular}[c]{@{}c@{}}Drone-Mic\\ Distance\end{tabular}} &
  \multicolumn{1}{c|}{Sound Source} &
  \multicolumn{1}{c|}{\begin{tabular}[c]{@{}c@{}}Drone-Source\\ Distance\end{tabular}} &
  \multicolumn{1}{c|}{\begin{tabular}[c]{@{}c@{}}Source\\ Loudness\end{tabular}} &
  \multicolumn{1}{c|}{Room} &
  \begin{tabular}[c]{@{}c@{}}Recording\\ Duration\end{tabular} \\ \hline \hline
\multicolumn{9}{|c|}{\textit{Drone Noise + Sound Source Recordings}} \\ \hline
\multicolumn{1}{|c|}{\multirow{2}{*}{\dronelarge}} &
  \multicolumn{1}{c|}{\multirow{4}{*}{\begin{tabular}[c]{@{}c@{}}Low,\\ High\end{tabular}}} &
  \multicolumn{1}{c|}{\multirow{4}{*}{\begin{tabular}[c]{@{}c@{}}\micup,\\ \micdown,\\ \miccenter\end{tabular}}} &
  \multicolumn{1}{c|}{\multirow{4}{*}{\begin{tabular}[c]{@{}c@{}}25cm,\\ 50cm\end{tabular}}} &
  \multicolumn{1}{c|}{\multirow{4}{*}{\begin{tabular}[c]{@{}c@{}}human vocal sounds,\\ non-vocal human-\\ presence,\\ ambient sounds\end{tabular}}} &
  \multicolumn{1}{c|}{1/3/5 m} &
  \multicolumn{1}{c|}{\multirow{2}{*}{\begin{tabular}[c]{@{}c@{}}60 dB,\\ 90 dB\end{tabular}}} &
  \multicolumn{1}{c|}{$\text{room}_1$} & 5 hours
   \\ \cline{6-6} \cline{8-9} 
\multicolumn{1}{|c|}{} &
  \multicolumn{1}{c|}{} &
  \multicolumn{1}{c|}{} &
  \multicolumn{1}{c|}{} &
  \multicolumn{1}{c|}{} &
  \multicolumn{1}{c|}{3/6/9 m} &
  \multicolumn{1}{c|}{} &
  \multicolumn{1}{c|}{$\text{room}_2$} & 5 hours
   \\ \cline{1-1} \cline{6-9} 
\multicolumn{1}{|c|}{\multirow{2}{*}{\dronesmall}} &
  \multicolumn{1}{c|}{} &
  \multicolumn{1}{c|}{} &
  \multicolumn{1}{c|}{} &
  \multicolumn{1}{c|}{} &
  \multicolumn{1}{c|}{1/3/5 m} &
  \multicolumn{1}{c|}{\multirow{2}{*}{60 dB}} &
  \multicolumn{1}{c|}{$\text{room}_1$} & 2.5 hours
   \\ \cline{6-6} \cline{8-9} 
\multicolumn{1}{|c|}{} &
  \multicolumn{1}{c|}{} &
  \multicolumn{1}{c|}{} &
  \multicolumn{1}{c|}{} &
  \multicolumn{1}{c|}{} &
  \multicolumn{1}{c|}{3/6/9 m} &
  \multicolumn{1}{c|}{} &
  \multicolumn{1}{c|}{$\text{room}_3$} & 2.5 hours
   \\ \hline
\multicolumn{9}{|c|}{\textit{Drone-Only Recordings}} \\ \hline
\multicolumn{1}{|c|}{\begin{tabular}[c]{@{}c@{}}\dronelarge,\\ \dronesmall\end{tabular}} &
  \multicolumn{1}{c|}{\begin{tabular}[c]{@{}c@{}}Low,\\ High\end{tabular}} &
  \multicolumn{1}{c|}{\begin{tabular}[c]{@{}c@{}}\micup,\\ \micdown,\\ \miccenter\end{tabular}} &
  \multicolumn{1}{c|}{\begin{tabular}[c]{@{}c@{}}25cm,\\ 50cm\end{tabular}} &
  \multicolumn{1}{c|}{-} &
  \multicolumn{1}{c|}{-} &
  \multicolumn{1}{c|}{-} &
  \multicolumn{1}{c|}{-} & 2.3 hours
   \\ \hline
\multicolumn{9}{|c|}{\textit{Source-Only Recordings}} \\ \hline
\multicolumn{1}{|c|}{\multirow{4}{*}{-}} &
  \multicolumn{1}{c|}{\multirow{4}{*}{-}} &
  \multicolumn{1}{c|}{\multirow{4}{*}{\begin{tabular}[c]{@{}c@{}}\micup,\\ \micdown,\\ \miccenter\end{tabular}}} &
  \multicolumn{1}{c|}{\multirow{4}{*}{\begin{tabular}[c]{@{}c@{}}25cm,\\ 50cm\end{tabular}}} &
  \multicolumn{1}{c|}{\multirow{4}{*}{\begin{tabular}[c]{@{}c@{}}human vocal sounds,\\ non-vocal human-\\ presence,\\ ambient sounds\end{tabular}}} &
  \multicolumn{1}{c|}{1/3/5 m} &
  \multicolumn{1}{c|}{\multirow{2}{*}{\begin{tabular}[c]{@{}c@{}}60 dB,\\ 90 dB\end{tabular}}} &
  \multicolumn{1}{c|}{$\text{room}_1$} & 2.5 hours
   \\ \cline{6-6} \cline{8-9} 
\multicolumn{1}{|c|}{} &
  \multicolumn{1}{c|}{} &
  \multicolumn{1}{c|}{} &
  \multicolumn{1}{c|}{} &
  \multicolumn{1}{c|}{} &
  \multicolumn{1}{c|}{3/6/9 m} &
  \multicolumn{1}{c|}{} &
  \multicolumn{1}{c|}{$\text{room}_2$} & 2.5 hours
   \\ \cline{6-9} 
\multicolumn{1}{|c|}{} &
  \multicolumn{1}{c|}{} &
  \multicolumn{1}{c|}{} &
  \multicolumn{1}{c|}{} &
  \multicolumn{1}{c|}{} &
  \multicolumn{1}{c|}{3/6/9 m} &
  \multicolumn{1}{c|}{60 dB} &
  \multicolumn{1}{c|}{$\text{room}_3$} & 1.2 hours
   \\ 
  \multicolumn{1}{|c|}{} &
  \multicolumn{1}{c|}{} &
  \multicolumn{1}{c|}{} &
  \multicolumn{1}{c|}{} &
  \multicolumn{1}{c|}{} &
  \multicolumn{1}{c|}{} &
  \multicolumn{1}{c|}{} &
  \multicolumn{1}{c|}{} & \\ \hline
 &
   &
   &
   &
   &
   &
   &
  \multicolumn{1}{c|}{} &
  \textbf{\recordingduration hours} \\ \cline{9-9} 
\end{tabular}}
\caption{Table summarizes all the data collection of \name, consisting of three recording settings -- \textit{combined drone and source}, \textit{drone-only} as well as \textit{source-only} recordings. Within each, we vary the attributes of -- drones (type, throttle), microphones (configuration, distance) and sources (signal type, distance, loudness). In total, our dataset amounts to \recordingduration hours of recording time.}
\label{tbl:data-collection-summary}
\vspace{-0.65cm}
\end{table}

\vspace{-0.1cm}
\subsection{Sound-Source Attributes}
\label{sec:dataset-sources}
\vspace{-0.2cm}

\noindent\textit{Source Signal Types.} We consider three categories of commonly occurring sounds in search and rescue settings: \textbf{human vocal (HV) sounds} such as speech, screams, and cries, compiled from Google’s Audioset \cite{gemmeke2017audioset}, Freesound\footnote{\url{https://freesound.org/}}, and audio distress dataset \cite{gaviria2020distress}, \textbf{human non-vocal (HNV) sounds} such as door knocks, and clapping, that indicate human presence, and ambient \textbf{non-human (NH) sounds}, such as fire crackling and water dripping, that may trigger false alarms. We curate HNV and NH categories from Freesound, with complete file details provided in Appendix Table~\ref{tbl:source-sounds}. The 15 hours of drone+source recordings listed in Table \ref{tbl:data-collection-summary} is split into HV, HNV, and NH in the ratio of 3:1:1, where HV category consists of equal parts of male, female, and baby crying sound types. %
These source audio files were played through a JBL Flip 6 Bluetooth speaker, with a frequency response from 63 Hz to 20 kHz, and a signal-to-noise ratio over 80 dB, sufficient to replicate low, mid and high frequencies of human speech, screams and non-verbal cues, as well as ambient sounds, such as, dripping water and burning fire. 

\noindent\textit{Source Loudness Types.} We calibrated audio playback at two intensity levels: typical scream loudness (90–100 dB) and typical speech loudness (60–70 dB) for \dronelarge, while only the quieter speech range for \dronesmall to evaluate performance under more challenging conditions.

\noindent\textit{Source Distances and Angles.} We perform experiments in three different indoor environments, -- a small conference room ($\text{room}_1$), two large multi-purpose halls ($\text{room}_2$ and $\text{room}_3$), within our university. In terms of dimensions, $\text{room}_1$ was a small sized room (6.5m $\times$ 5m $\times$ 2.5m), while $\text{room}_2$ (16.5m $\times$ 11m $\times$ 2.5m) and $\text{room}_3$ (14m $\times$ 9.5m $\times$ 6.9m). The three rooms have varying multi-path effects with reverberation times of 1.34s, 0.83s and 1.3s\footnote{Measured using: \url{https://www.soniflex.com/en/information-service/information-service/room-acoustics-measurement-app}}. In the small room (room$_1$), we chose source-to-drone horizontal distances 1m, 3m, and 5m, while in the big rooms (room$_2$ and room$_3$), we chose longer distances 3m, 6m, and 9m. Across all settings, we placed the speaker at a height of 40 cm above the ground on a stand, and an azimuth of $225^\circ$, with respect to the microphone array.

\begin{figure}[t!]
    \centering
    \includegraphics[width=1\linewidth]{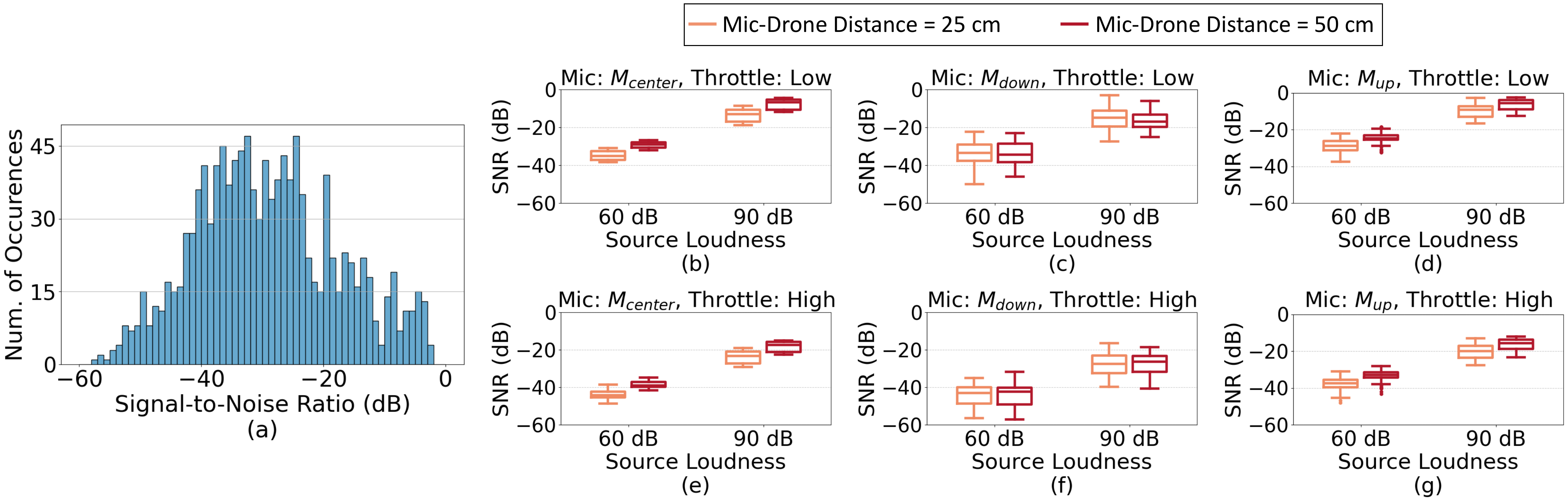}
    \caption{Figure (a) depicts the histogram of signal-to-noise ratios (SNRs) of all the data collected, and (b-g) depict the SNRs achieved across three different microphones and two drone throttle levels. Specifically, each box plot contrasts the SNR levels for different source loudness levels (60~dB and 90~dB), and microphone-drone distances (25~cm and 50~cm).}
    \label{fig:snr}
    \vspace{-0.5cm}
\end{figure}

\vspace{-0.1cm}
\subsection{Signal-to-Noise Ratio Distributions}
\label{sec:dataset-snr}
\vspace{-0.3cm}
We quantify dataset diversity using signal-to-noise ratios (SNRs), computed as \(SNR_{dB} = 20*log_{10}({RMS_{source}}/{RMS_{drone}})\), where $RMS_{source}$ and $RMS_{drone}$ represent root mean square amplitude values of individual source and drone recordings respectively. Figure~\ref{fig:snr}(a) visualizes the overall distribution of SNRs across all the different settings: drones and their throttles, microphones and their placements, source sounds and their volumes, and rooms. It shows that our dataset spans SNRs from -57.2 dB to -2.5 dB, with consistently negative values demonstrating drone noise dominance, a key challenge for noise suppression and human-presence detection. Detailed analysis in Figures~\ref{fig:snr}(b-g) shows expected patterns: higher source loudness (90 dB scream level) yields better SNRs than speech-level sounds (60 dB), while the microphones below the drone \micdown consistently underperforms compared to the other microphones due to wind noise exposure, exhibiting the lowest median SNR across comparable conditions of throttle level, source loudness, and mic-drone distance.

\vspace{-0.25cm}
\section{Evaluation}
\label{sec:evaluation}
\vspace{-0.2cm}
We evaluate two use-case scenarios of the \name dataset: (a) to detect human presence in indoor search and rescue, and (b) to derive design recommendations for drone-audition systems.%

\vspace{-0.2cm}
\subsection{Application 1: Detecting Human Presence}
\label{sec:eval-application1}
\vspace{-0.2cm}

\begin{figure}
    \centering
    \includegraphics[width=1\linewidth]{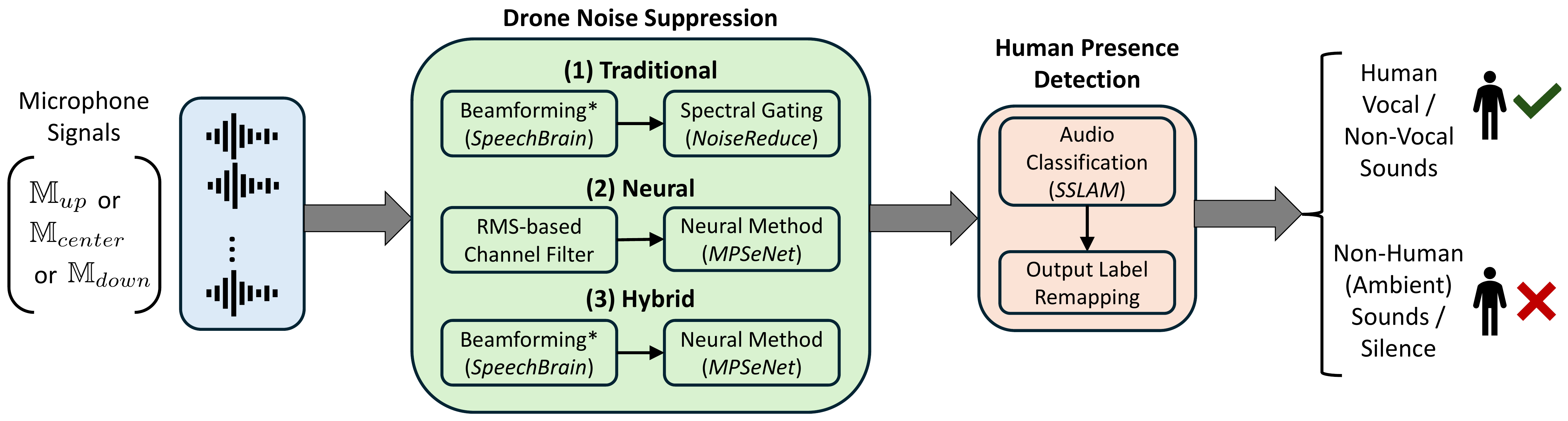}
    \caption{Figure illustrates the current benchmarking pipeline, that takes as input the audio recordings from one of the microphones, \micup, \miccenter or \micdown. Our pipeline consists of drone noise suppression (which we evaluate for three combinations of algorithms), followed by human-presence detection, that ultimately predicts audio file as corresponding to human sounds (vocal, non-vocal) or non-human sounds (ambient), or silence. Here $^*$ indicates the absence of beamforming for microphone, \miccenter, as it consists of only a single microphone.
    }
    \label{fig:pipeline}
\end{figure}

We outline a two-stage system for a drone-audition system aimed at detecting human presence: 1) \textit{Noise Suppression}, to remove the noise generated or caused by the drone, and enhance the sounds indicating human presence, and 2) \textit{Identification of Human Presence} through classification of the enhanced audio from the first stage as either human or non-human. %
A detailed block diagram of these stages is provided in Figure \ref{fig:pipeline}.

\noindent\textbf{Noise Suppression Baselines and Metrics.} We evaluate the effectiveness of the following state-of-the-art audio enhancement methods on the collected drone audio dataset:

\textbullet~\textit{Traditional Enhancement} \textbf{(Trad.)}.  Given that we have 8-channel microphone arrays, \micdown and \micup, we perform the well-known Minimum Variance Distortionless Response (MVDR) beamforming\footnote{For the standalone single-channel microphone, \miccenter, we skip beamforming as it is not applicable.} using the SpeechBrain library~\cite{souden2009optimal, speechbrain, speechbrain_v1}. MVDR takes as inputs the \textit{source's location}, i.e., (x,y,z) co-ordinates to calculate the azimuth and elevation direction, %
as well as the \textit{drone-only} signals, for enhancing directional signals corresponding to the source while suppressing the drone noise. Subsequently, %
we perform noise suppression through \textit{spectral gating} using NoiseReduce library~\cite{sainburg2020finding, tim_sainburg_2019_3243139}, which gates all noise below their corresponding frequency-specific thresholds, that are estimated for every short-time window, to account for non-stationary signals such as drone noise. We set the following values to the parameters: \texttt{stationary}:\texttt{False} and \texttt{aggressiveness}:\texttt{0.5}. 

\textbullet~\textit{Neural Enhancement} \textbf{(Neural)}. We leverage MPSENet (Magnitude and Phase Speech Enhancement Network)~\cite{lu2023mp}, which is a neural enhancement technique designed for suppressing noise while preserving speech integrity, due to its accurate magnitude and phase estimation. It has proven effective for the VoiceBank+DEMAND as well as Deep Noise Suppression Challenge datasets and has an open-source model. 
As MPSENET can only take a single audio channel as input, we choose the channel with least root mean square (RMS) amplitude among the 8-ch microphones, assuming it would have the least amount of drone noise.

\textbullet~\textit{Beamforming + Neural Enhancement} \textbf{(Hybrid)}. Here, similar to the neural approach, we leverage MPSeNet noise suppression, but instead of choosing a channel based on least RMS, we utilize the beamformed signal as input for improved noise robustness.

\noindent{\textit{Metrics.}} We use the Scale-Invariant Signal-to-Distortion Ratio (\textbf{SI-SDR}) metric which measures the quality of a reconstructed signal by comparing it to a reference signal while being invariant to scale changes -- \(\text{SI-SDR} = 10\log_{10}\left({|\alpha\mathbf{x}|^2}/{|\alpha\mathbf{x}-\hat{\mathbf{x}}|^2}\right)\), where \(\alpha = {\hat{\mathbf{x}}^\intercal\mathbf{x}}/{|\mathbf{x}|^2}\),  
$\mathbf{x}$ is the reference signal and $\hat{\mathbf{x}}$ is the estimate. We prefer SI-SDR over other metrics such as PESQ due to its suitability for non-speech sounds. SI-SDR is computed for each recording configuration (in Table~\ref{tbl:data-collection-summary}), and mean and standard deviation are computed across all audio segments within an SNR group (informed by Figure~\ref{fig:snr}).

\noindent\textbf{Audio Event Detection Baselines and Metrics.}
Here, our goal is to classify the noise-suppressed, drone-infused source recordings into three classes: vocal sounds, non-vocal human sounds and ambient non-human sounds, to ultimately detect human presence. For this purpose, we adopt \textbf{SSLAM} (Self-Supervised Learning from Audio Mixtures) \cite{alex2025sslam}, a state-of-the-art audio transformer pre-trained on AudioSet~\cite{gemmeke2017audioset} as an audio classifier to test on our dataset. SSLAM is designed to robustly detect and classify source sound in presence of other overlapping sounds, and is shown to outperform other existing models in the task of audio classification on audioset and has an open source model. %
Pre-training on AudioSet also ensures coverage of diverse acoustic events. 

\noindent{\textit{Metrics.}} We map SSLAM's 527 AudioSet output labels into four audio classes: human vocal (HV), human non-vocal (HNV), non-human (NH), as well as \textit{silence}, using deterministic rules (e.g., "Speech" → HV, "Clap"→ HNV, "Wind" → NH). We provide the lookup table in our project's GitHub repository. Subsequently, we use \textbf{per-class F1-score}, which is the harmonic mean of per-class precision and recall, to report the classification performance of the model for the three source classes. Similar to SI-SDR, F1-score is computed across all audio segments per configuration, and mean and standard deviation are computed across all F1-scores of segments %
within an SNR group. 

\noindent{\textbf{Results:}}
\noindent\textbf{Noise Suppression.}
Table \ref{tab:combined_results} shows the SI-SDR scores of the three noise suppression methods (\textbf{Trad.}, \textbf{Neural}, \textbf{Hybrid}) as well as the original unprocessed audio (\textbf{No Proc.}) in four SNR groups. As shown in Figure~\ref{fig:snr}(b-g), low SNRs typically result from high drone throttle or low source volume, while high SNRs arise from the opposite conditions.

\begin{table}[ht]
\centering
\small
\vspace{-0.3cm}
\resizebox{\textwidth}{!}{\begin{tabular}{
    l
    l
    S[table-format=-2.2]
    @{${}\pm{}$}
    S[table-format=1.2]
    S[table-format=-2.2]
    @{${}\pm{}$}
    S[table-format=1.2]
    S[table-format=-2.2]
    @{${}\pm{}$}
    S[table-format=1.2]
    S[table-format=1.3]
    @{${}\pm{}$}
    S[table-format=1.3]
    S[table-format=1.3]
    @{${}\pm{}$}
    S[table-format=1.3]
    S[table-format=1.3]
    @{${}\pm{}$}
    S[table-format=1.3]
}
\toprule
\multirow{2}{*}{\textbf{SNR Group}} & \multirow{2}{*}{\textbf{Method}} & 
\multicolumn{6}{c}{\textbf{SI-SDR (dB)}} & 
\multicolumn{6}{c}{\textbf{F1 Score}} \\
\cmidrule(lr){3-8} \cmidrule(lr){9-14}
 & & \multicolumn{2}{c}{HV} & \multicolumn{2}{c}{HNV} & \multicolumn{2}{c}{NH} & \multicolumn{2}{c}{HV} & \multicolumn{2}{c}{HNV} & \multicolumn{2}{c}{NH} \\
\midrule
\hline

\multirow{4}{*}{$>-10$ dB} 
 & No Proc. & -17.03 & 6.97 & -21.94 & 5.75 & -11.86 & 4.75 &  \multicolumn{2}{c}{-} & \multicolumn{2}{c}{-} & \multicolumn{2}{c}{-}  \\
 & Trad. & -14.12 & 7.32 & -17.75 & 7.05 & -10.60 & 5.35 &\multicolumn{2}{c}{-} & \multicolumn{2}{c}{-} & \multicolumn{2}{c}{-}  \\
 & Neural & \textbf{-12.37} & 8.72 & \textbf{-14.03} & 8.35 & -12.33 & 9.24 & \textbf{0.874} & 0.007 & 0.232 & 0.025 & \textbf{0.192} & 0.078 \\
 & Hybrid & -14.00 & 8.36 & -14.94 & 7.77 & -15.32 & 8.89 & 0.829 & 0.011 & \textbf{0.280} & 0.076 & 0.000 & 0.000 \\
\midrule

\multirow{4}{*}{-20 to -10 dB} 
 & No Proc. & -22.74 & 5.57 & -25.91 & 4.05 & -18.41 & 5.56 &\multicolumn{2}{c}{-} & \multicolumn{2}{c}{-} & \multicolumn{2}{c}{-} \\
 & Trad. & -21.19 & 6.00 & -24.27 & 4.72 & -18.42 & 5.72 &\multicolumn{2}{c}{-} & \multicolumn{2}{c}{-} & \multicolumn{2}{c}{-}  \\
 & Neural & \textbf{-14.30} & 8.76 & \textbf{-16.33} & 8.44 & \textbf{-15.80} & 9.48 & \textbf{0.843} & 0.055 & \textbf{0.190} & 0.076 & \textbf{0.274} & 0.057 \\
 & Hybrid & -16.05 & 8.08 & -17.52 & 7.60 & -18.33 & 7.57 & 0.755 & 0.095 & 0.108 & 0.109 & 0.020 & 0.024 \\
\midrule

\multirow{4}{*}{-30 to -20 dB} 
 & No Proc. & -28.38 & 2.72 & -29.03 & 1.74 & -26.43 & 3.82 &  \multicolumn{2}{c}{-} & \multicolumn{2}{c}{-} & \multicolumn{2}{c}{-}  \\
 & Trad. & -28.26 & 2.44 & -28.95 & 1.67 & -27.27 & 3.67 &\multicolumn{2}{c}{-} & \multicolumn{2}{c}{-} & \multicolumn{2}{c}{-}  \\
 & Neural & \textbf{-19.77} & 7.07 & -22.18 & 5.89 & \textbf{-20.81} & 6.34 & \textbf{0.423} & 0.173 & \textbf{0.072} & 0.025 & \textbf{0.169} & 0.127 \\
 & Hybrid & -20.25 & 6.11 & \textbf{-21.70} & 5.05 & -21.30 & 5.48 & 0.344 & 0.066 & 0.011 & 0.019 & 0.050 & 0.030 \\
\midrule

\multirow{4}{*}{$<-30$ dB} 
 & No Proc. & -29.50 & 1.70 & -29.41 & 1.57 & -29.08 & 2.04 &\multicolumn{2}{c}{-} & \multicolumn{2}{c}{-} & \multicolumn{2}{c}{-} \\
 & Trad. & -29.54 & 1.62 & -29.38 & 1.43 & -29.42 & 2.10 &\multicolumn{2}{c}{-} & \multicolumn{2}{c}{-} & \multicolumn{2}{c}{-}  \\
 & Neural & -23.44 & 5.28 & -23.70 & 4.65 & -23.50 & 5.12 & 0.255 & 0.071 & \textbf{0.043} & 0.030 & \textbf{0.130} & 0.030 \\
 & Hybrid & \textbf{-22.76} & 5.26 & \textbf{-22.77} & 4.76 & \textbf{-22.85} & 5.00 & \textbf{0.311} & 0.142 & 0.025 & 0.023 & 0.067 & 0.054 \\
\midrule

\multicolumn{2}{c}{Clean Recordings (No drone)} 
 & \multicolumn{2}{c}{-} & \multicolumn{2}{c}{-} & \multicolumn{2}{c}{-} & \multicolumn{2}{c}{0.968} & \multicolumn{2}{c}{0.852} & \multicolumn{2}{c}{0.796} \\
\bottomrule
\end{tabular}}
\vspace{0.1cm}
\caption{Comparison of noise suppression performance (SI-SDR in dB) and classification accuracy (F1 scores) across different SNR groups. Traditional \cite{sainburg2020finding, souden2009optimal}, Neural Net \cite{lu2023mp}, and Hybrid methods are shown with mean $\pm$ std values computed across audio files within each SNR group.%
}
\label{tab:combined_results}
\vspace{-0.5cm}
\end{table}

Noise suppression performance degrades across all methods as SNR decreases, with neural approaches (Neural and Hybrid) consistently outperforming traditional method, especially in extreme low-SNR conditions (<-20 dB). For HV sounds, neural methods achieve the largest gains (e.g., -12.37 dB vs. -17.03 dB at SNR >-10 dB), as existing techniques are optimized for vocal patterns. However, HNV and NH sounds prove challenging, with marginal improvements even for hybrid methods, as these algorithms have not been trained for such sounds. The hybrid approach achieves similar performance to neural alone in high SNR sounds, while at lower SNRs, the hybrid approach tends to perform better, possibly due to preserving the signal from the source direction with beamforming. %
However, accurately finding the source direction automatically at low SNR conditions would be challenging. Notably, all methods struggle when SNR falls below -30 dB, highlighting the need for advanced solutions in extreme noisy drone environments beyond conventional speech enhancement methods.

\noindent\textbf{Results: Classification.} 
As shown in Table \ref{tab:combined_results}, we compute the F1-score for classification of the noise suppressed audio from the two best noise suppression methods (Neural and Hybrid), %
as well as the original clean (or drone noise-free) recordings. %
HV sounds showed better classification performance (F1=0.87 ± 0.007 for SNR>-10 dB) compared to HNV and NH sounds, approaching clean recording performance at high SNR. A large number of HNV and NH instances got classified as silence. This disparity stems from the noise suppression stage, where these methods achieved better SI-SDR improvements for vocal sounds enabling cleaner reconstruction, resulting in more accurate downstream classification. Classification performance improved with higher input SNR across all categories, mirroring noise suppression results where low SNR reduced SI-SDR gains. This direct link between suppression quality and classification accuracy highlights the need for end-to-end pipeline optimization in drone audition systems. Refer to Figures \ref{fig:app1noisesupp} and \ref{fig:app1classification} in Appendix for further visualization of the trends of noise suppression and classification performances with varying SNRs.

\vspace{-0.3cm}
\subsection{Application 2: Deriving Design Recommendations for Drone-Audition Systems}
\label{sec:application2}
\vspace{-0.2cm}
The \name dataset enables hardware design recommendations for drone-audition systems through recordings with varied parameter configurations. %
By analyzing acoustic performance across these parameters of the dataset and the noise suppression analysis (Section \ref{sec:eval-application1}), we derive practical guidelines and highlight key trade-offs\footnote{For statistical significance of comparisons in this section, we applied Shapiro-Wilk normality testing, followed by Tukey HSD (if normal distribution) or Wilcoxon signed-rank with Bonferroni correction (if non-normal)}.%

\begin{figure}[htbp]
\vspace{-0.3cm}
    \centering
    \begin{minipage}[b]{0.48\textwidth}
    \centering
\includegraphics[width=1.05\linewidth]%
{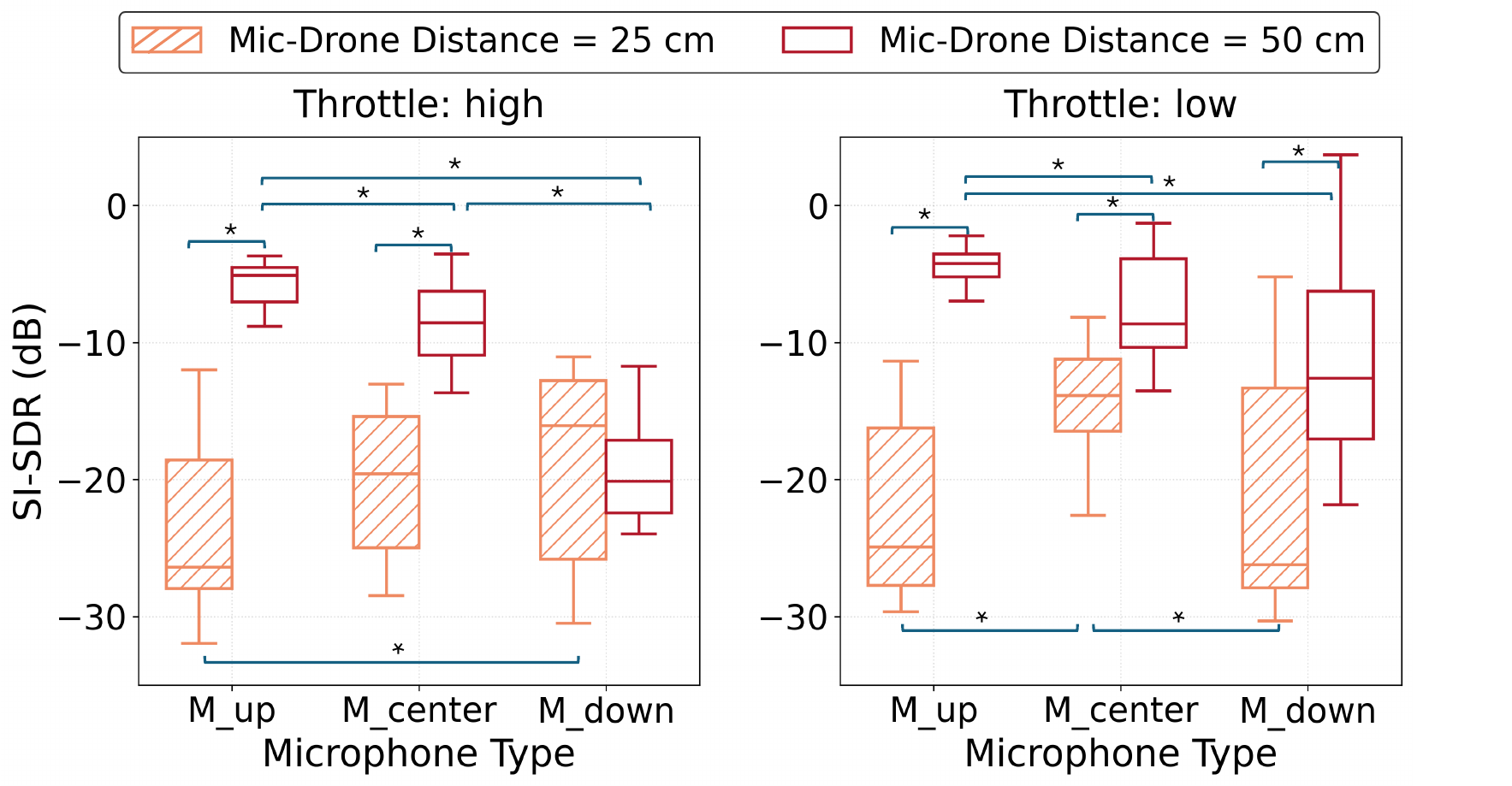}
    \caption{SI-SDR scores for \textit{Hybrid} noise suppression on HV sounds, for varied microphone types, drone-microphone distances, at (a) high, (b) low throttles. * indicate statistically significant difference$^{11}$ between pairs ($p<0.05$).} 
    \label{fig:app2mic}
    \end{minipage}
    \hfill
    \begin{minipage}[b]{0.48\textwidth}
        \centering\includegraphics[width=\linewidth]%
        {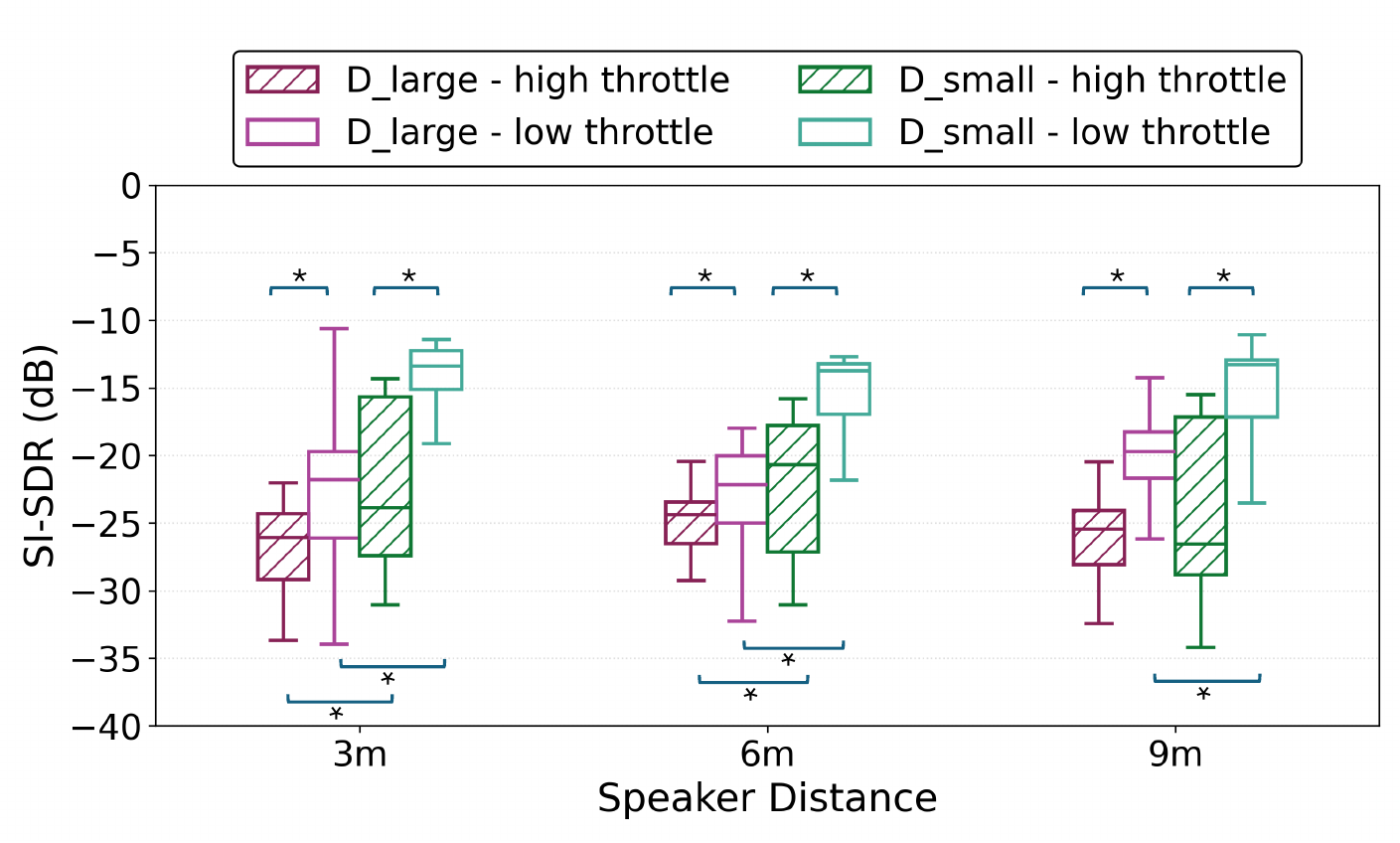}
    \caption{SI-SDR scores for \textit{Hybrid} noise suppression on HV sounds, for two drones, at two throttles, as well as drone-speaker distances. * indicate statistically significant difference$^{11}$ between pairs ($p<0.05$).}
    \label{fig:app2drone}
    \end{minipage}
    \vspace{-0.3cm}
\end{figure}
\noindent\textbf{Microphone Placement/Distance Trade-off}:
In general, below-drone microphone (\micdown) achieves the least SI-SDR score due to direct wind noise exposure from propellors, compared to the above-drone microphones (\micup, \miccenter). %
However, below-drone placement offers proximity to ground-level sound sources. At drone-microphone distance of 25~cm and high throttle setting (shown in Figure~\ref{fig:app2mic}(a)\footnote{In Figure~\ref{fig:app2mic}, we consider a higher SNR setting: source vol: 90 dB, \dronelarge, $\text{room}_1$, drone-source distance: 1m}), all microphones perform poorly, but increasing the distance to 50~cm significantly improves performance for all settings of above-drone microphones as well as low throttle setting of below-drone microphone (see Figures~\ref{fig:app2mic}(a) and \ref{fig:app2mic}(b)). %
Overall, these findings suggest that while above-drone placements are preferable due to their higher SI-SDR scores, below-drone placements may be viable at lower throttle settings with greater suspension distance  (see $\S$\ref{sec:discussion}). 
%These findings suggest that when microphones are required to be mounted closer to the drone body (e.g., for balance), above-drone placements are preferable, while below-drone placement may be viable if greater suspension distance is possible (see $\S$\ref{sec:discussion}). 

\noindent\textbf{Microphone Array Trade-offs}:
The 8-channel array above the drone (\micup) provides beamforming capabilities, resulting in significant improvement in the SI-SDR (p<0.05) over a single-channel microphone (\miccenter) in high SNR conditions, e.g., low throttle, 50 cm drone-to-microphone distance (Fig. \ref{fig:app2mic}(b)). However, microphone array has higher processing demands, hence system designers must weigh these trade-offs against mission needs. While multi-microphone arrays are valuable for precise source localization, single-microphone setups may be preferable for missions where memory and power constraints are critical. 

\noindent\textbf{Drone Throttle Adjustments}:
As depicted in Figure \ref{fig:app2drone}, at a low SNR setting\footnote{In Figure~\ref{fig:app2drone}, we consider a lower SNR setting: source vol: 60 dB, drone-microphone distance: 50 cm}, the SI-SDR performance improves significantly (p<0.05) at lower throttle across both drones, \dronelarge and \dronesmall, compared to high throttle. 
This suggests that drone audition systems should incorporate adaptive throttle reduction strategies during critical listening periods, particularly when detecting faint sounds or during search patterns where audio detection outweighs mobility needs.

\noindent\textbf{Drone Size Trade-offs}:
Figure \ref{fig:app2drone} shows that \dronesmall achieves higher SI-SDR scores than \dronelarge at both low and high throttle levels, even at higher drone-source distances, demonstrating the difference in ego-noise across drones. In general, larger drones (>1.5~kg payload) support advanced recording setups, compared to smaller drones that enable only lightweight configurations. Hence, mission planning must weigh payload capacity against acoustic performance requirements.

\vspace{-0.3cm}
\section{Discussion}
\label{sec:discussion}
\vspace{-0.3cm}
\noindent \textit{New Research Opportunities.} 
The \name dataset %
provides a publicly available benchmarking resource for developing next-generation noise suppression algorithms and audio classification models capable of operating under the extreme low-SNR conditions characteristic of drone recordings. 
Researchers can leverage the dataset's controlled variations in throttle levels, and microphone configurations, to train models that adapt to different noise profiles. Our dataset includes separately recorded source and drone sounds that would support data augmentation strategies for training as well as for simulating multi-source conditions. %
While the recordings were made at a single azimuth, the controlled variations in the dataset could be combined with spatial augmentation techniques to simulate multi-azimuth scenarios. Thus, beyond human presence detection, our dataset %
offers potential utility for sound localization, speech recovery and other applications in mobile robotics.

\noindent \textit{Broader Impact. }
This work enables positive societal impact through improved search-and-rescue capabilities, particularly in disaster scenarios where visual systems fail. However, we acknowledge potential negative applications in unauthorized surveillance, as enhanced drone audition could compromise privacy if misused. While our dataset focuses on humanitarian applications, we recommend deployment safeguards including access controls for sensitive data, onboard processing to minimize raw audio transmission, and clear usage guidelines for ethical adoption.%

\noindent \textit{Limitations and Future Work.}
While our dataset spans 216 unique configurations, including variations in drones, microphones, and recording environments, it does not encompass the whole distribution of acoustic profiles encountered in real-world scenarios. We perform experiments across three distinct indoor environments, additionally varying the drone-to-source distance to introduce diversity in reverberation and multipath effects. However, we acknowledge that this alone is insufficient to represent the broad spectrum of cluttered, reflective, and absorptive materials that may be present in real-world search and rescue contexts. Data augmentation approaches based on synthetic room impulse responses, such as those proposed by Tang et al.~\cite{tang2022gwa}, offer a promising direction for expanding the dataset and enhancing model generalization under reverberant conditions. 
Moreover, while this study primarily focused on detecting human distress sounds for rescue purposes, future work could extend to the detection of non-human auditory cues relevant to emergency response, such as the acoustic signatures of fire, structural collapse, to provide a more comprehensive situational awareness framework.

Another limitation stems from our method of simulating drone hovering by attaching the drone to a frame. Although this setup emulates a hovering drone, it does not fully capture the micro-dynamics of real hovering, such as the subtle balancing adjustments that can influence the drone's acoustic profile. However, our data-collection design enables controlled measurements across a wide range of drone-microphone and drone-source distances, provides recommendations for developing the recording setup in a real-world hovering scenario. Future work should include practical deployment considerations such as suspending microphones versus mounting microphones close to the body of the drone, payload capacity of the drone, as well as computational considerations of on-board versus off-board audio processing. %

In addition, our current dataset primarily represents indoor and semi-controlled acoustic conditions. Outdoor environments differ substantially, with dominant sources of wind noise rather than reverberation or multipath reflections, and typically lower signal-to-noise ratios due to larger drone–source distances. Expanding the dataset to include outdoor recordings would therefore add valuable diversity and realism. An incremental next step could involve conducting low-risk outdoor flights and synthetically augmenting the captured data with target sounds to simulate realistic rescue scenarios.

While a more diverse and realistic dataset would be beneficial, we believe that \name is a significant initial step towards enabling audio capture in drone-based search and rescue.

\vspace{-0.3cm}
\section{Conclusion}
\vspace{-0.3cm}
We present \name dataset which is designed for human presence detection in drone-based indoor search and rescue settings. The dataset includes acoustic profiles in a wide range of SNR values from -57.2 dB to -2.5 dB, collected with various drones, microphones and source sounds. Overall, our dataset accounts to \recordingduration hours of recording duration, making it the largest drone audition dataset to the best of our knowledge. We also benchmark SOTA noise suppression and audio event detection methods, which work poorly, especially in the presence of non-vocal human sounds and ambient sounds. Moving forward, we hope our dataset serves as a stepping stone for enabling drone audition, even beyond search and rescue settings.  

\vspace{-0.3cm}
\section{Acknowledgments}
\vspace{-0.3cm}
We would like to thank the research assistants, Nihirra Kakkar and Ho Him Chan for their help during the recording sessions. %This research is supported by Ministry of Defence Singapore for the collaborative project between CNRS, France and NUS, Singapore called ``AMANIA : Augmented huMAN-swarm InterAction'', under the sub-project Blind Drone (TDP for MINDEF-DGA Joint Lab, Sub-Project Number: 9023200454-02).

\bibliographystyle{plain}
\bibliography{paper}

\begin{thebibliography}{10}

\bibitem{fixed-wing-vs-quadcopter-2}
Propeller Aero.
\newblock Fixed-wing \& vtol drones vs quadcopters for surveying: How to decide.
\newblock \url{https://www.propelleraero.com/blog/fixed-wing-vtol-drones-vs-quadcopters-for-surveying-how-to-decide/}, Accessed 2025.

\bibitem{alex2025sslam}
Tony Alex, Sara Atito, Armin Mustafa, Muhammad Awais, and Philip J~B Jackson.
\newblock {SSLAM}: Enhancing self-supervised models with audio mixtures for polyphonic soundscapes.
\newblock In {\em The Thirteenth International Conference on Learning Representations}, 2025.

\bibitem{dji-f450}
Amazon.
\newblock Dji f450 drone kit.
\newblock \url{https://www.amazon.sg/HAWKS-WORK-Quadcopter-Brushless-Transmitter/dp/B0DG2L1TQL/}, Accessed 2025.

\bibitem{deleforge2020drone}
Antoine Deleforge.
\newblock Drone audition for search and rescue: Datasets and challenges.
\newblock In {\em QUIET DRONES International Symposium on UAV/UAS Noise}, 2020.

\bibitem{dji-f330}
DJI.
\newblock Dji f330 drone kit.
\newblock \url{https://www-v1.dji.com/flame-wheel-arf/spec.html}, Accessed 2025.

\bibitem{fixed-wing-vs-quadcopter-1}
DroneNerds.
\newblock Fixed wing vs. quadcopter: Which to use when?
\newblock \url{https://enterprise.dronenerds.com/blog/enterprise-drone-program/fixed-wing-vs-quadcopter-which-to-use-when/}, Accessed 2025.

\bibitem{dubey2023icassp}
Harishchandra Dubey, Ashkan Aazami, Vishak Gopal, Babak Naderi, Sebastian Braun, Ross Cutler, Hannes Gamper, Mehrsa Golestaneh, and Robert Aichner.
\newblock Icassp 2023 deep noise suppression challenge.
\newblock In {\em ICASSP}, 2023.

\bibitem{fonseca2022FSD50K}
Eduardo Fonseca, Xavier Favory, Jordi Pons, Frederic Font, and Xavier Serra.
\newblock {FSD50K}: an open dataset of human-labeled sound events.
\newblock {\em IEEE/ACM Transactions on Audio, Speech, and Language Processing}, 30:829--852, 2022.

\bibitem{gaviria2020distress}
Jorge~Felipe Gaviria, Alejandra Escalante-Perez, Juan~Camilo Castiblanco, Nicolas Vergara, Valentina Parra-Garces, Juan~David Serrano, Andres~Felipe Zambrano, and Luis~Felipe Giraldo.
\newblock Deep learning-based portable device for audio distress signal recognition in urban areas.
\newblock {\em Applied Sciences}, 10(21), 2020.

\bibitem{gebru2021datasheets}
Timnit Gebru, Jamie Morgenstern, Briana Vecchione, Jennifer~Wortman Vaughan, Hanna Wallach, Hal~Daum{\'e} Iii, and Kate Crawford.
\newblock Datasheets for datasets.
\newblock {\em Communications of the ACM}, 64(12):86--92, 2021.

\bibitem{gemmeke2017audioset}
Jort~F. Gemmeke, Daniel P.~W. Ellis, Dylan Freedman, Aren Jansen, Wade Lawrence, R.~Channing Moore, Manoj Plakal, and Marvin Ritter.
\newblock Audio set: An ontology and human-labeled dataset for audio events.
\newblock In {\em 2017 IEEE International Conference on Acoustics, Speech and Signal Processing (ICASSP)}, pages 776--780, 2017.

\bibitem{kolamunna2021droneprint}
Harini Kolamunna, Thilini Dahanayaka, Junye Li, Suranga Seneviratne, Kanchana Thilakaratne, Albert~Y. Zomaya, and Aruna Seneviratne.
\newblock Droneprint: Acoustic signatures for open-set drone detection and identification with online data.
\newblock 5(1), March 2021.

\bibitem{lu2023mp}
Ye-Xin Lu, Yang Ai, and Zhen-Hua Ling.
\newblock Mp-senet: A speech enhancement model with parallel denoising of magnitude and phase spectra.
\newblock {\em arXiv preprint arXiv:2305.13686}, 2023.

\bibitem{martinez2020review}
Jose Martinez-Carranza and Caleb Rascon.
\newblock A review on auditory perception for unmanned aerial vehicles.
\newblock {\em Sensors}, 20(24), 2020.

\bibitem{morito2016partially}
Takayuki Morito, Osamu Sugiyama, Ryosuke Kojima, and Kazuhiro Nakadai.
\newblock Partially shared deep neural network in sound source separation and identification using a uav-embedded microphone array.
\newblock In {\em 2016 IEEE/RSJ International Conference on Intelligent Robots and Systems (IROS)}, pages 1299--1304. IEEE, 2016.

\bibitem{morito2016reduction}
Takayuki Morito, Osamu Sugiyama, Satoshi Uemura, Ryosuke Kojima, and Kazuhiro Nakadai.
\newblock Reduction of computational cost using two-stage deep neural network for training for denoising and sound source identification.
\newblock In {\em Trends in Applied Knowledge-Based Systems and Data Science: 29th International Conference on Industrial Engineering and Other Applications of Applied Intelligent Systems, IEA/AIE 2016, Morioka, Japan, August 2-4, 2016, Proceedings 29}, pages 562--573. Springer, 2016.

\bibitem{panayotov2015librispeech}
Vassil Panayotov, Guoguo Chen, Daniel Povey, and Sanjeev Khudanpur.
\newblock Librispeech: An asr corpus based on public domain audio books.
\newblock In {\em 2015 IEEE International Conference on Acoustics, Speech and Signal Processing (ICASSP)}, pages 5206--5210, 2015.

\bibitem{politis2022starss22}
Archontis Politis, Kazuki Shimada, Parthasaarathy Sudarsanam, Sharath Adavanne, Daniel Krause, Yuichiro Koyama, Naoya Takahashi, Shusuke Takahashi, Yuki Mitsufuji, and Tuomas Virtanen.
\newblock Starss22: A dataset of spatial recordings of real scenes with spatiotemporal annotations of sound events.
\newblock {\em arXiv preprint arXiv:2206.01948}, 2022.

\bibitem{speechbrain_v1}
Mirco Ravanelli, Titouan Parcollet, Adel Moumen, Sylvain de~Langen, Cem Subakan, Peter Plantinga, Yingzhi Wang, Pooneh Mousavi, Luca~Della Libera, Artem Ploujnikov, Francesco Paissan, Davide Borra, Salah Zaiem, Zeyu Zhao, Shucong Zhang, Georgios Karakasidis, Sung-Lin Yeh, Pierre Champion, Aku Rouhe, Rudolf Braun, Florian Mai, Juan Zuluaga-Gomez, Seyed~Mahed Mousavi, Andreas Nautsch, Ha~Nguyen, Xuechen Liu, Sangeet Sagar, Jarod Duret, Salima Mdhaffar, Ga{{\"e}}lle Laperri{{\`e}}re, Mickael Rouvier, Renato~De Mori, and Yannick Est{{\`e}}ve.
\newblock Open-source conversational ai with speechbrain 1.0.
\newblock {\em Journal of Machine Learning Research}, 25(333), 2024.

\bibitem{speechbrain}
Mirco Ravanelli, Titouan Parcollet, Peter Plantinga, Aku Rouhe, Samuele Cornell, Loren Lugosch, Cem Subakan, Nauman Dawalatabad, Abdelwahab Heba, Jianyuan Zhong, Ju-Chieh Chou, Sung-Lin Yeh, Szu-Wei Fu, Chien-Feng Liao, Elena Rastorgueva, François Grondin, William Aris, Hwidong Na, Yan Gao, Renato~De Mori, and Yoshua Bengio.
\newblock {SpeechBrain}: A general-purpose speech toolkit, 2021.
\newblock arXiv:2106.04624.

\bibitem{reddy2020interspeech}
CK~Reddy, E~Beyrami, H~Dubey, V~Gopal, R~Cheng, R~Cutler, S~Matusevych, R~Aichner, A~Aazami, S~Braun, et~al.
\newblock The interspeech 2020 deep noise suppression challenge: Datasets, subjective speech quality and testing framework. arxiv 2020.
\newblock {\em arXiv preprint arXiv:2001.08662}, 2020.

\bibitem{ruiz2018aira}
Oscar Ruiz-Espitia, Jose Martinez-Carranza, and Caleb Rascon.
\newblock Aira-uas: an evaluation corpus for audio processing in unmanned aerial system.
\newblock In {\em 2018 International Conference on Unmanned Aircraft Systems (ICUAS)}, pages 836--845. IEEE, 2018.

\bibitem{tim_sainburg_2019_3243139}
Tim Sainburg.
\newblock timsainb/noisereduce: v1.0, June 2019.

\bibitem{sainburg2020finding}
Tim Sainburg, Marvin Thielk, and Timothy~Q Gentner.
\newblock Finding, visualizing, and quantifying latent structure across diverse animal vocal repertoires.
\newblock {\em PLoS computational biology}, 16(10):e1008228, 2020.

\bibitem{serizel2020sound}
Romain Serizel, Nicolas Turpault, Ankit Shah, and Justin Salamon.
\newblock Sound event detection in synthetic domestic environments.
\newblock In {\em ICASSP 2020-2020 IEEE International Conference on Acoustics, Speech and Signal Processing (ICASSP)}, pages 86--90. IEEE, 2020.

\bibitem{souden2009optimal}
Mehrez Souden, Jacob Benesty, and Sofiene Affes.
\newblock On optimal frequency-domain multichannel linear filtering for noise reduction.
\newblock {\em IEEE Transactions on audio, speech, and language processing}, 18(2):260--276, 2009.

\bibitem{strauss2018dregon}
Martin Strauss, Pol Mordel, Victor Miguet, and Antoine Deleforge.
\newblock Dregon: Dataset and methods for uav-embedded sound source localization.
\newblock In {\em 2018 IEEE/RSJ International Conference on Intelligent Robots and Systems (IROS)}, pages 1--8. IEEE, 2018.

\bibitem{tang2022gwa}
Zhenyu Tang, Rohith Aralikatti, Anton~Jeran Ratnarajah, and Dinesh Manocha.
\newblock Gwa: A large high-quality acoustic dataset for audio processing.
\newblock In {\em ACM SIGGRAPH 2022 Conference Proceedings}, pages 1--9, 2022.

\bibitem{turpault2019sound}
Nicolas Turpault, Romain Serizel, Ankit~Parag Shah, and Justin Salamon.
\newblock Sound event detection in domestic environments with weakly labeled data and soundscape synthesis.
\newblock In {\em Workshop on Detection and Classification of Acoustic Scenes and Events}, 2019.

\bibitem{wang2019audio}
Lin Wang, Ricardo Sanchez-Matilla, and Andrea Cavallaro.
\newblock Audio-visual sensing from a quadcopter: dataset and baselines for source localization and sound enhancement.
\newblock In {\em 2019 IEEE/RSJ International Conference on Intelligent Robots and Systems (IROS)}, pages 5320--5325. IEEE, 2019.

\bibitem{yano2024ground}
Tsubasa Yano, Benjamin Yen, Katsutoshi Itoyama, and Kazuhiro Nakadai.
\newblock Ground surface material classification with drone noise.
\newblock 2024.

\end{thebibliography}

\newpage
\section*{NeurIPS Paper Checklist}

\begin{enumerate}

\item {\bf Claims}
    \item[] Question: Do the main claims made in the abstract and introduction accurately reflect the paper's contributions and scope?
    \item[] Answer: \answerYes{} %
    \item[] Justification: The dataset \name size, characteristics, and usage scenarios have been summarized both in the abstract and the last paragraph of introduction section.

\item {\bf Limitations}
    \item[] Question: Does the paper discuss the limitations of the work performed by the authors?
    \item[] Answer: \answerYes{} %
    \item[] Justification: Section 5 has a subsection on \textit{Limitations and Future Work}.

\item {\bf Theory assumptions and proofs}
    \item[] Question: For each theoretical result, does the paper provide the full set of assumptions and a complete (and correct) proof?
    \item[] Answer: \answerNA{} %
    \item[] Justification: The paper does not have theoretical results.

    \item {\bf Experimental result reproducibility}
    \item[] Question: Does the paper fully disclose all the information needed to reproduce the main experimental results of the paper to the extent that it affects the main claims and/or conclusions of the paper (regardless of whether the code and data are provided or not)?
    \item[] Answer: \answerYes{} %
    \item[] Justification: The details of our dataset collection setup has been detailed in Section 3, other supporting details are in Appendix, the dataset has been made publicly available.

\item {\bf Open access to data and code}
    \item[] Question: Does the paper provide open access to the data and code, with sufficient instructions to faithfully reproduce the main experimental results, as described in supplemental material?
    \item[] Answer: \answerYes{} %
    \item[] Justification: We release the dataset publicly, and the link is provided in the abstract. For benchmarking, we use code from published papers that have their code released, which we provide references to in the evaluation section. 

\item {\bf Experimental setting/details}
    \item[] Question: Does the paper specify all the training and test details (e.g., data splits, hyperparameters, how they were chosen, type of optimizer, etc.) necessary to understand the results?
    \item[] Answer: \answerYes{} %
    \item[] Justification: While we do not train a model in this paper, wherever appropriate, we mention the hyperparameters used for inference in Section~\ref{sec:evaluation}. %

\item {\bf Experiment statistical significance}
    \item[] Question: Does the paper report error bars suitably and correctly defined or other appropriate information about the statistical significance of the experiments?
    \item[] Answer: \answerYes{} %
    \item[] Justification: Provided in Table \ref{tab:combined_results}, Figures \ref{fig:app2mic} and \ref{fig:app2drone} in the main paper, and Figures \ref{fig:app1noisesupp} and \ref{fig:app1classification} in the Appendix.

\item {\bf Experiments compute resources}
    \item[] Question: For each experiment, does the paper provide sufficient information on the computer resources (type of compute workers, memory, time of execution) needed to reproduce the experiments?
    \item[] Answer: \answerYes{} %
    \item[] Justification: We mention about the computation time and resources needed to run the benchmarking experiments in Appendix. 
    
\item {\bf Code of ethics}
    \item[] Question: Does the research conducted in the paper conform, in every respect, with the NeurIPS Code of Ethics \url{https://neurips.cc/public/EthicsGuidelines}?
    \item[] Answer: \answerYes{} %
    \item[] Justification: The dataset did not involve collecting or storing information or data from people. It only contains drone recordings and audio recordings from publicly available datasets. The dataset is being released under MIT license.

\item {\bf Broader impacts}
    \item[] Question: Does the paper discuss both potential positive societal impacts and negative societal impacts of the work performed?
    \item[] Answer: \answerYes{} %
    \item[] Justification: The positive and negative societal impact of this dataset is covered in the Discussion section ($\S$\ref{sec:discussion}).
    
\item {\bf Safeguards}
    \item[] Question: Does the paper describe safeguards that have been put in place for responsible release of data or models that have a high risk for misuse (e.g., pretrained language models, image generators, or scraped datasets)?
    \item[] Answer: \answerYes{} %
    \item[] Justification: The source sound files representing human speech or cries for help, and other environmental sounds have been carefully curated from public datasets and online resources, thus ensuring that no sensitive information is present in these recordings. The rest of the dataset is recordings of these source sound files along with drone noise. Thus it does not pose a high risk of misuse. We have added this briefly in the Discussion section ($\S$\ref{sec:discussion}).

\item {\bf Licenses for existing assets}
    \item[] Question: Are the creators or original owners of assets (e.g., code, data, models), used in the paper, properly credited and are the license and terms of use explicitly mentioned and properly respected?
    \item[] Answer: \answerYes{} %
    \item[] Justification: All the source audio files are curated from publicly available datasets~\cite{panayotov2015librispeech,gaviria2020distress, gemmeke2017audioset} for research purposes, which have been referenced in the Dataset section ($\S$\ref{sec:dataset}). In particular, for the sounds taken from \textit{FreeSound}, we include links to the individual audio files in the Appendix. Likewise, we cite all the existing models~\cite{sainburg2020finding, tim_sainburg_2019_3243139, speechbrain, speechbrain_v1, lu2023mp} used for our benchmarking in our Evaluation section ($\S$\ref{sec:evaluation}). 

\item {\bf New assets}
    \item[] Question: Are new assets introduced in the paper well documented and is the documentation provided alongside the assets?
    \item[] Answer: \answerYes{} %
    \item[] Justification: Following the guidelines from Gebru \textit{et al.}~\cite{gebru2021datasheets}, we include a datasheet for our dataset in the Appendix, which compliments our  publicly released dataset.

\item {\bf Crowdsourcing and research with human subjects}
    \item[] Question: For crowdsourcing experiments and research with human subjects, does the paper include the full text of instructions given to participants and screenshots, if applicable, as well as details about compensation (if any)? 
    \item[] Answer: \answerNA{} %
    \item[] Justification: \answerNA{}

\item {\bf Institutional review board (IRB) approvals or equivalent for research with human subjects}
    \item[] Question: Does the paper describe potential risks incurred by study participants, whether such risks were disclosed to the subjects, and whether Institutional Review Board (IRB) approvals (or an equivalent approval/review based on the requirements of your country or institution) were obtained?
    \item[] Answer: \answerNA{} %
    \item[] Justification: \answerNA{}

\item {\bf Declaration of LLM usage}
    \item[] Question: Does the paper describe the usage of LLMs if it is an important, original, or non-standard component of the core methods in this research? Note that if the LLM is used only for writing, editing, or formatting purposes and does not impact the core methodology, scientific rigorousness, or originality of the research, declaration is not required.
    \item[] Answer: \answerNA{} %
    \item[] Justification: \answerNA{}

\end{enumerate}

\appendix
\newpage

% \section*{Appendix for \name}
% We provide additional information about data collection and evaluation, as well as \name Datasheet in the following sections.

% For a quick glimpse, check out our \textbf{Demo Webpage}: 

% \url{https://apps.ahlab.org/DroneAudioSet-code/}

% and \textbf{Github Repo}: 

% \url{https://github.com/augmented-human-lab/DroneAudioSet-code.git}.
\section{Additional Information about Data Collection}
\label{app:dataset-additional-info}
In Section~\ref{sec:dataset}, we provide details on the different data collection configurations. Here, in Section~\ref{app:supplementary-dataset-info}, we provide additional supporting information -- on the drones, source sounds, as well as indoors environments where experiments were performed. Subsequently, in Sections~\ref{app:analysis-drone-noise-profile} and \ref{app:analysis-source-profile}, we provide detailed analysis on our recorded drone noise as well as source audio profiles, respectively.

\subsection{Data Collection Configurations}
\label{app:supplementary-dataset-info}
\begin{figure}[h!]
    \centering
    \includegraphics[width=0.9\linewidth]{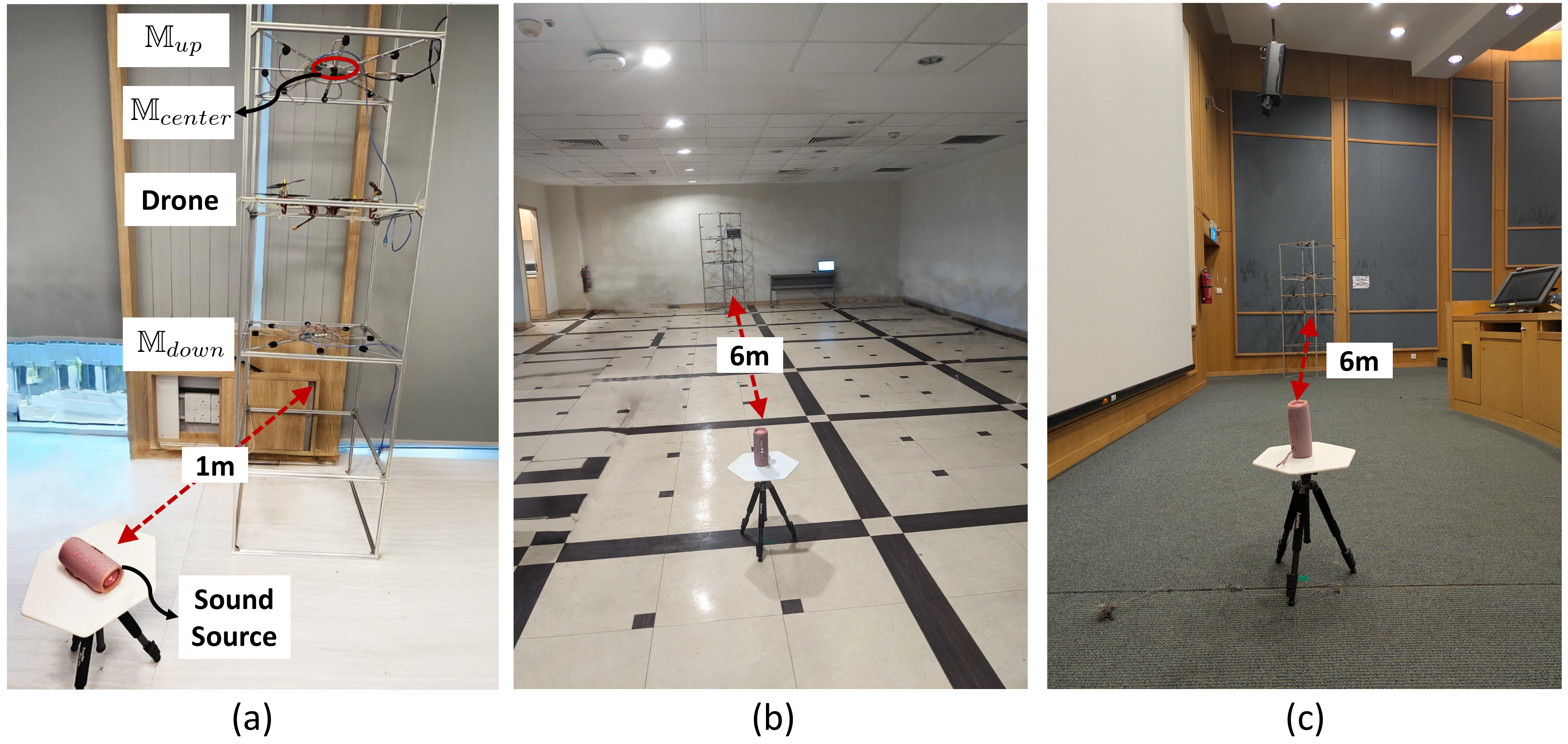}
    \caption{Figure depicts \name's recording setup in the three environments -- (a) $\text{room}_1$, (b) $\text{room}_2$, and (c) $\text{room}_3$, with the drone-source distance of 1~m, 6~m and 6~m, respectively. (a) additionally annotates the Bluetooth speaker used as sound source, the drone affixed to the frame, as well as the three microphone arrays, \micup, \miccenter, and \micdown. }
    \label{fig:recordingsetup}
\end{figure}

Figure~\ref{fig:recordingsetup}(a-c) depicts the three indoor environments, a small conference room ($\text{room}_1$), as well as two large multi-purpose halls ($\text{room}_2$, $\text{room}_3$), along with the drone recording setup. Table~\ref{tbl:drone-specs} provides detailed specifications of the drones, including their model, size, weight, as well as maximum flight times. In particular, the differences in frame weights as well as the propeller sizes of the two drones results in variations in the noise produced (see $\S$\ref{app:analysis-drone-noise-profile}), adding diversity to our dataset. Table~\ref{tbl:drone-throttle} depicts the maximum sound pressure (SPL) measured at a 1~m distance from the drones, when operated at the `low' and `high' throttles. As shown, we set the throttles such that the SPL levels of the two drones, \dronelarge and \dronesmall, are comparable, i.e., within 4~dBA of each other, at both throttle settings. Finally, in Table~\ref{tbl:source-sounds}, we enumerate all the sound sources we gather from public repositories online, for the different source types, along with their licenses.     

\begin{table}[h!]
\centering
\renewcommand{\arraystretch}{1}
    \resizebox{\textwidth}{!}{
    \begin{tabular}{|l|l|l|}
\hline
Drone Name          & \dronelarge             & \dronesmall            \\ \hline \hline
Model               & DJI F450 Quadcopter~\cite{dji-f450} & DJI F330 Quadcopter~\cite{dji-f330}        \\ \hline
Frame Diagonal Wheelbase    & 450 mm  &  330 mm \\ \hline
Frame Weight & 282g & 156g \\\hline
Max Take-off Weight & 1.8 kg              & 1.2 kg              \\ \hline
Max Flight Time     & 20 minutes          & 20 minutes         \\ \hline
Propellers & Diameter = 9.4 inch, Pitch = 5.0 inch & Diameter = 8 inch, Pitch =  4.5 inch \\ \hline
\end{tabular}}

\caption{Comparison of the specifications of the two quadcopter drones considered for our data collection.}%
\label{tbl:drone-specs}
\end{table}

\begin{table}[h!]
\centering
\begin{tabular}{|l|l|l|}
\hline
Drone/Throttle & Throttle=Low & Throttle=High \\ \hline\hline
\dronelarge        & 76.9 dBA     & 91.8 dBA      \\ \hline
\dronesmall        & 74.9 dBA     & 88.3 dBA      \\ \hline
\end{tabular}
\caption{Table enumerating the Sound Pressure Levels or SPLs (measured in the dBA scale) to quantify the drone throttle levels. }
\label{tbl:drone-throttle}
\end{table}

\begin{table}[]
\renewcommand{\arraystretch}{1}
    \resizebox{0.8\textwidth}{!}{
\begin{tabular}{|ll|}
\hline
\multicolumn{1}{|l|}{Source Type} &
  \multicolumn{1}{l|}{References} \\ \hline \hline
\multicolumn{2}{|c|}{\textbf{Source Type: Human Sounds}} \\ \hline
\multicolumn{1}{|l|}{Male Speech/Scream} & \begin{tabular}[c]{@{}l@{}}Librispeech \cite{panayotov2015librispeech} (CC BY 4.0)\\
   Audio Distress Dataset \cite{gaviria2020distress} (CC BY 4.0)\\
   Google's Audioset \cite{gemmeke2017audioset} (CC BY 4.0)\\
   https://freesound.org/people/tarrei/sounds/141242/\\
   https://freesound.org/people/bevibeldesign/sounds/316648/\\
   https://freesound.org/people/Feed\_/sounds/529115/\\
   https://freesound.org/people/Sgt.Dukenberry/sounds/696683/\\
   https://freesound.org/people/guamorims/sounds/391365/
   \end{tabular}\\
   \hline
\multicolumn{1}{|l|}{Female Speech/Scream} &
   \begin{tabular}[c]{@{}l@{}}Librispeech \cite{panayotov2015librispeech} (CC BY 4.0)\\
   Audio Distress Dataset \cite{gaviria2020distress} (CC BY 4.0)\\
   Google's Audioset \cite{gemmeke2017audioset} (CC BY 4.0)\\
   https://freesound.org/people/pyro13djt/sounds/338811/\\
   https://freesound.org/people/AmeAngelofSin/sounds/326893/\\
   https://freesound.org/people/Deathstardude/sounds/360851/\\
   https://freesound.org/people/megmcduffee/sounds/393488/\\
   https://freesound.org/people/marionagm90/sounds/220663/
   
   \end{tabular}\\
   \hline
\multicolumn{1}{|l|}{Baby Cries} &\begin{tabular}[c]{@{}l@{}}Google's Audioset \cite{gemmeke2017audioset} (CC BY 4.0)\\
https://freesound.org/people/josephvm/sounds/442655/\\
https://freesound.org/people/gumballworld/sounds/398552/\end{tabular}

   \\ \hline
\multicolumn{2}{|c|}{\textbf{Source Type: Non-Vocal Human Sounds}} \\ \hline
\multicolumn{1}{|l|}{Clapping} &
  \begin{tabular}[c]{@{}l@{}}https://freesound.org/people/soundsofscienceupf/sounds/463500/\\ https://freesound.org/people/MinecraftM153/sounds/634110/\\ https://freesound.org/people/parkersenk/sounds/452709/\\ https://freesound.org/people/Vrymaa/sounds/734612/\\ https://freesound.org/people/RanneM/sounds/475559/\end{tabular} \\ \hline
\multicolumn{1}{|l|}{Door Knocking} &
  \begin{tabular}[c]{@{}l@{}}https://freesound.org/people/deleted\_user\_7146007/sounds/383756/\\ https://freesound.org/people/Terry93D/sounds/342549/\\ https://freesound.org/people/TRP/sounds/573835/\end{tabular} \\ \hline
\multicolumn{1}{|l|}{Finger Snapping} &
  https://freesound.org/people/shutup\_outcast/sounds/367836/ \\ \hline
\multicolumn{1}{|l|}{Footsteps} &
  \begin{tabular}[c]{@{}l@{}}https://freesound.org/people/craigsmith/sounds/480592/\\ https://freesound.org/people/florianreichelt/sounds/456037/\\ https://freesound.org/people/taure/sounds/363470/\end{tabular} \\ \hline
\multicolumn{1}{|l|}{Glass Knocking} &
  \begin{tabular}[c]{@{}l@{}}https://freesound.org/people/launemax/sounds/249932/\\ https://freesound.org/people/frisko28i/sounds/339135/\end{tabular} \\ \hline
\multicolumn{1}{|l|}{Metal Banging} &
  \begin{tabular}[c]{@{}l@{}}https://freesound.org/people/MarcelWagner/sounds/639658/\\ https://freesound.org/people/bruno.auzet/sounds/661640/\\ https://freesound.org/people/BenjaminNelan/sounds/410362/\\ https://freesound.org/people/KaBlazik\_Samples/sounds/561006/\end{tabular} \\ \hline
\multicolumn{2}{|c|}{\textbf{Source Type: Non-Human Sounds}} \\ \hline
\multicolumn{1}{|l|}{Alarm Ringing} &
  \begin{tabular}[c]{@{}l@{}}https://freesound.org/people/Tempouser/sounds/123349/\\ https://freesound.org/people/Gerent/sounds/558727/\\ https://freesound.org/people/columbia23/sounds/397097/\\ https://freesound.org/people/SoundBiterSFX/sounds/730939/\\ https://freesound.org/people/OmegaZeta/sounds/112211/\end{tabular} \\ \hline
\multicolumn{1}{|l|}{Fire Burning} &
  \begin{tabular}[c]{@{}l@{}}https://freesound.org/people/craigsmith/sounds/675763/\\ https://freesound.org/people/ken788/sounds/386751/\\ https://freesound.org/people/bruno.auzet/sounds/528079/\\ https://freesound.org/people/Soonus/sounds/528662/\\ https://freesound.org/people/raremess/sounds/222558/\end{tabular} \\ \hline
\multicolumn{1}{|l|}{Object Dropping} &
  \begin{tabular}[c]{@{}l@{}}https://freesound.org/people/Nox\_Sound/sounds/556263/\\ https://freesound.org/people/moxobna/sounds/131230/\\ https://freesound.org/people/Starvolt/sounds/189216/\\ https://freesound.org/people/ursenfuns/sounds/440595/\\ https://freesound.org/people/dr19/sounds/427090/\end{tabular} \\ \hline
\multicolumn{1}{|l|}{Traffic Honking} &
  \begin{tabular}[c]{@{}l@{}}https://freesound.org/people/wanaki/sounds/569613/\\ https://freesound.org/people/DeVern/sounds/349922/\\ https://freesound.org/people/danlucaz/sounds/517673/\\ https://freesound.org/people/specrad1/sounds/571317/\\ https://freesound.org/people/allthingssound/sounds/424918/\end{tabular} \\ \hline
\multicolumn{1}{|l|}{Vacuum Cleaner Humming} &
  \begin{tabular}[c]{@{}l@{}}https://freesound.org/people/SaintOche/sounds/706109/\\ https://freesound.org/people/Exacom/sounds/273194/\\ https://freesound.org/people/IlseHimschoot/sounds/495547/\\ https://freesound.org/people/SamuelGremaud/sounds/511736/\\ https://freesound.org/people/Huminaatio/sounds/159348/\end{tabular} \\ \hline
\multicolumn{1}{|l|}{Water Dripping} &
  \begin{tabular}[c]{@{}l@{}}https://freesound.org/people/TRP/sounds/616854/\\ https://freesound.org/people/ani\_music/sounds/632456/\\ https://freesound.org/people/florianreichelt/sounds/451761/\\ https://freesound.org/people/wakey/sounds/211390/\\ https://freesound.org/people/FairSonicStudio/sounds/530624/\end{tabular} \\ \hline
\end{tabular}}
\caption{Table depicts the different types of source sounds curated for creating vocal human sounds, non-vocal human sounds as well as non-human (ambient) sounds for \name dataset. Here, all sounds from \texttt{FreeSound} repository have the Creative Commons license. }
\label{tbl:source-sounds}
\end{table}

\begin{figure}[ht!]
    \centering
    \includegraphics[width=1\linewidth]{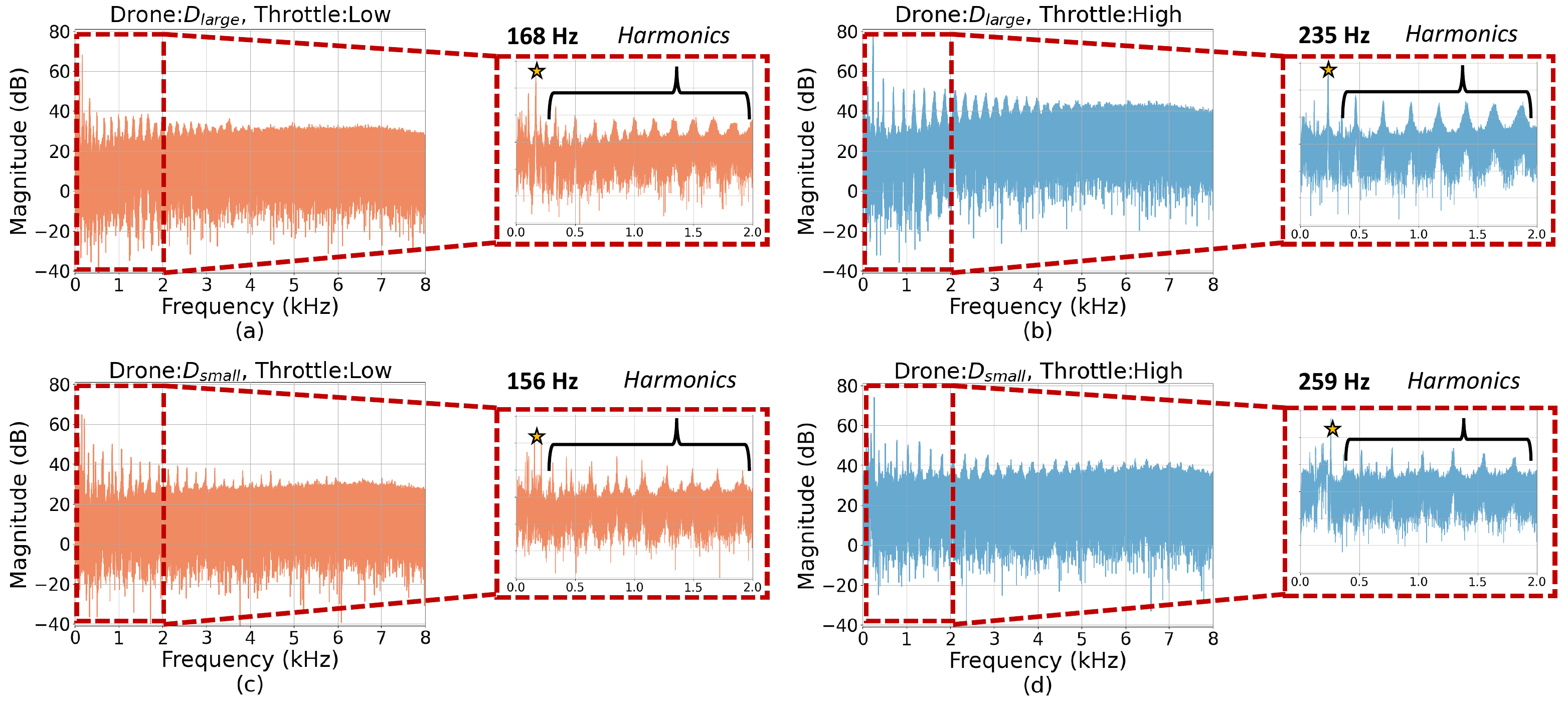}
    \caption{Figure depicts the frequency spectrum of the drones, \dronelarge at -- (a) low throttle and (b) high throttle, as well as \dronesmall at -- (c) low throttle and (d) high throttle.}
    \label{fig:drone-throttle-analysis}
\end{figure}

\begin{figure}[ht!]
    \centering
    \includegraphics[width=1\linewidth]{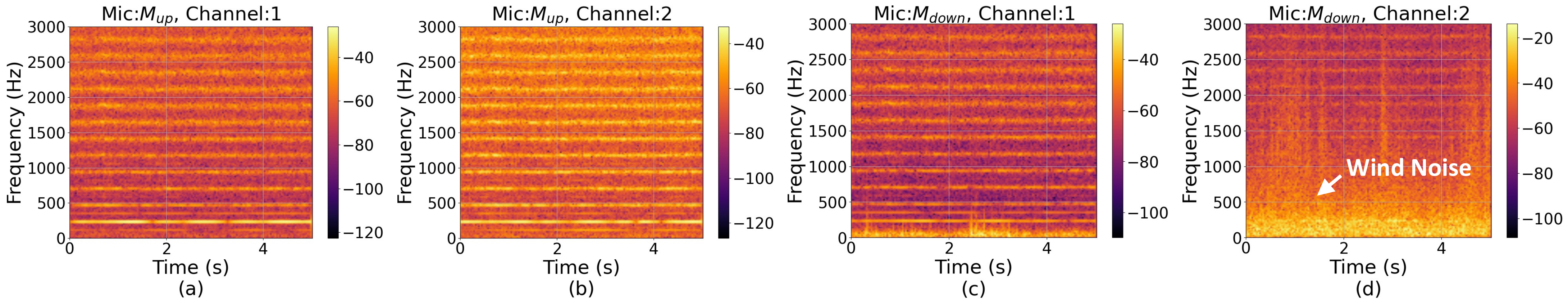}
    \caption{Figure depicts the spectrogram of two microphones, \micup at -- (a) channel 1 and (b) channel 2, as well as \micdown at -- (c) channel 1 and (d) channel 2. Here, the microphone array's channel 1 represents a microphone location that is in-between two propellers, while channel 2 represents a location directly above or below the propeller for \micup and \micdown, respectively.}
    \label{fig:drone-mic-analysis}
\end{figure}

\subsection{Drone Noise Profiles}
\label{app:analysis-drone-noise-profile}
Drones have unique audio characteristics due to the presence of motors and propellers, each of which contributes to the tonal (i.e., pure tones) and broadband sounds (i.e., energy in wider frequency bands), respectively. Two specific factors that contribute to variations in a drone's sounds are -- (a) throttle level, which controls the amount of power provided to the motors, and (b) the microphone placement, around the drone. Below, we discuss these two aspects for the drone, \dronelarge. %

\noindent{\textit{Effect of Throttle.}} Figures~\ref{fig:drone-throttle-analysis} (a-d) depict the magnitude spectrum of both the drone sounds, captured from the microphone, \micup, at two different throttle levels, notated as \textit{low} and \textit{high}, respectively. From the figures, we observe that the drone sounds have a certain tonal sound, with a fundamental frequency (denoted by`$\star$'), as well as harmonics, which are its multiples. Furthermore, we observe that higher throttle results in higher fundamental frequency, as expected due to increase in the motor's number of rotations per minute (RPM). In addition to the tonal aspect, we observe that in both cases, the drone sounds have significant energy up to 8~kHz, which is due to the turbulent airflow contributed by the movement of the propeller blades. As expected, higher throttle has higher broadband noise levels (exceeding 40~dB from the figure) due to faster blade rotation speeds.     

\noindent{\textit{Effect of Microphone Location.}} Due to the turbulent airflow around the drone, the location of the microphone can significantly affect the captured sound. Figure~\ref{fig:drone-mic-analysis} depicts the variations in the sounds captured by two channels of \micup and \micdown microphone arrays. As shown, while the tonal aspects remain similar across them all, the low-frequency wind noise is significant in the \micdown, especially in the channel that is right under the propeller blade. This increase in noise can be attributed to the thrust generated by drone's propulsion which pushes air in the downward direction, thereby increasing the wind noise. On the flip side, microphone \micdown can be suitable due to its proximity to sound sources, thereby justifying our choice for recording data from different placements of microphone arrays.   

\begin{figure}[t!]
    \centering
    \includegraphics[width=1\linewidth]{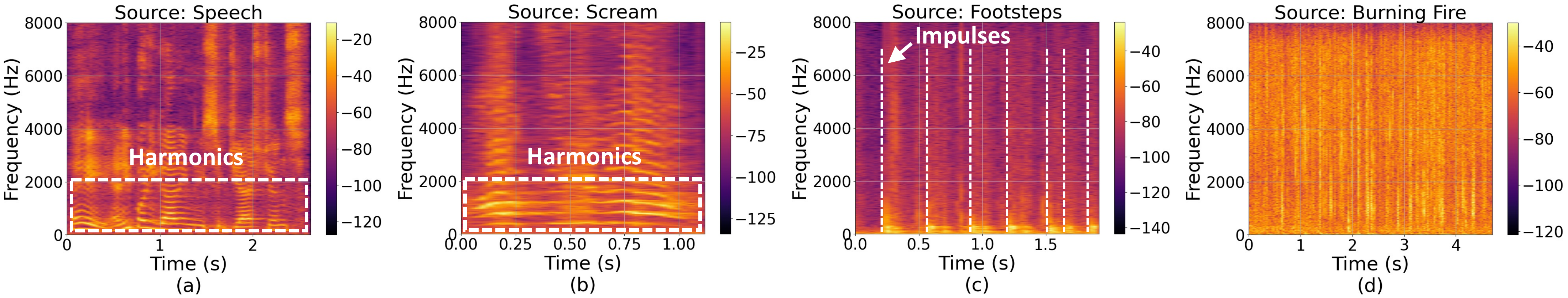}
    \caption{Frequency spectrum of different types of source sounds, namely three human-presence indicating sounds such as speech, screams and footsteps, as well as a non-human presence sound of burning fire.}
    \label{fig:source-fft}
\end{figure}

\subsection{Source Audio Profiles}
\label{app:analysis-source-profile}
As discussed earlier in Section~\ref{sec:dataset-sources}, we utilize three categories of sounds of interest, namely, human vocal sounds, non-vocal human-presence sounds, as well as non-human, ambient sounds. In Figure~\ref{fig:source-fft}, we demonstrate spectrograms of speech and scream sounds (human vocal), footsteps (human non-vocal), as well as burning fire sounds (ambient) to demonstrate the distinct time-frequency patterns of the different categories. In particular, we observe that the human vocal sounds have significant energy in lower frequencies, with harmonics below 4~kHz, and scream signals having higher fundamental frequency or pitch compared to speech signals. The non-vocal human sounds, such as footsteps and knocking, have transient and impulse-like pattern, creating bursts of high energy. On the other hand, ambient sounds such as vacuum cleaner sounds and burning fire have broadband distribution, with uniform energy across all frequencies, similar to gaussian noise. Given that the drone sounds have tonal and broadband patterns up to 8~kHz, they significantly interfere with all the above source signals, making drone noise suppression challenging.

\section{Additional Information about Evaluation}
% \subsection{Demo Webpage}
% Readers are encouraged to visit our demo webpage: \url{https://apps.ahlab.org/DroneAudioSet-code/} to listen to some sample recordings and the noise suppressed outputs from the different techniques.

% \subsection{Label mapping}
% The 527 classes of Audioset were mapped to the 4 classes HV (human vocals), HNV (human non-vocals), NH (non-human), and ND (not detected) sounds according to this mapping table: \url{https://github.com/augmented-human-lab/DroneAudioSet-code/blob/main/audioset_labelmapping.csv}. Only the class \textit{silence} was mapped to ND.

\subsection{Additional figures for Application 1: Detecting Human Presence}
\begin{figure}[htbp]
    \centering
    \begin{minipage}[b]{1\textwidth}
    \centering
\includegraphics[width=\linewidth]{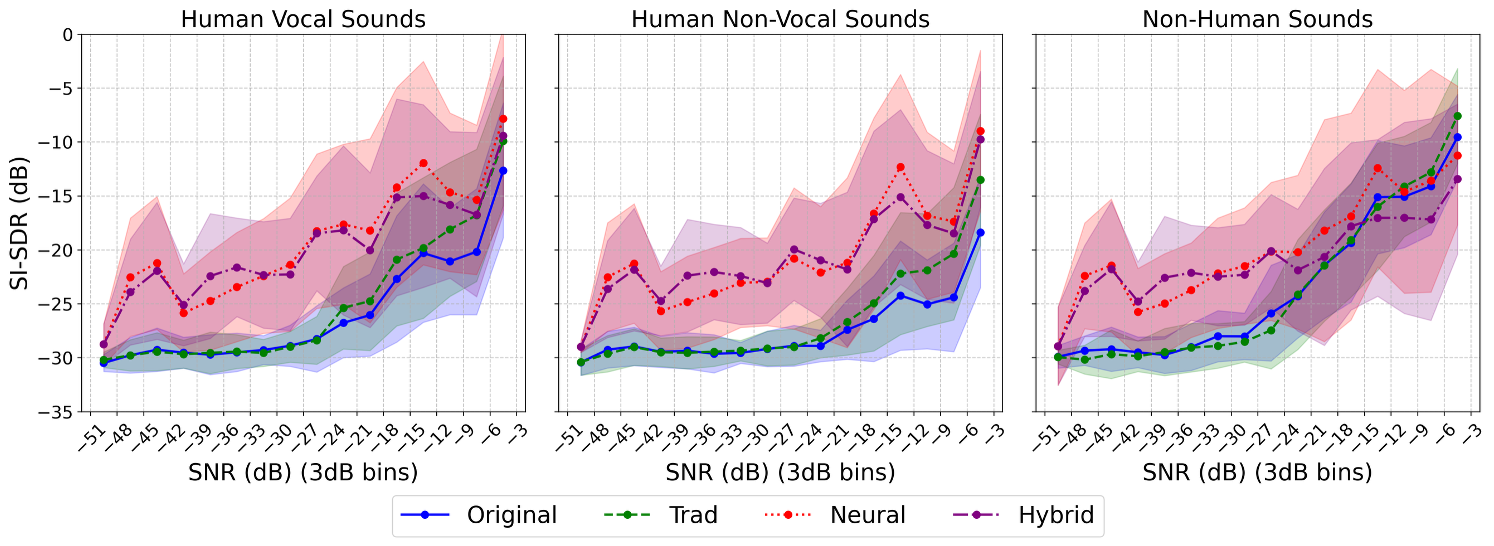}
    \caption{Performance (SI-SDR) of noise suppression techniques on recordings in \name at different SNRs.}
    \label{fig:app1noisesupp}
    \end{minipage}
    \hfill
    \begin{minipage}[b]{1\textwidth}
        \centering\includegraphics[width=\linewidth]{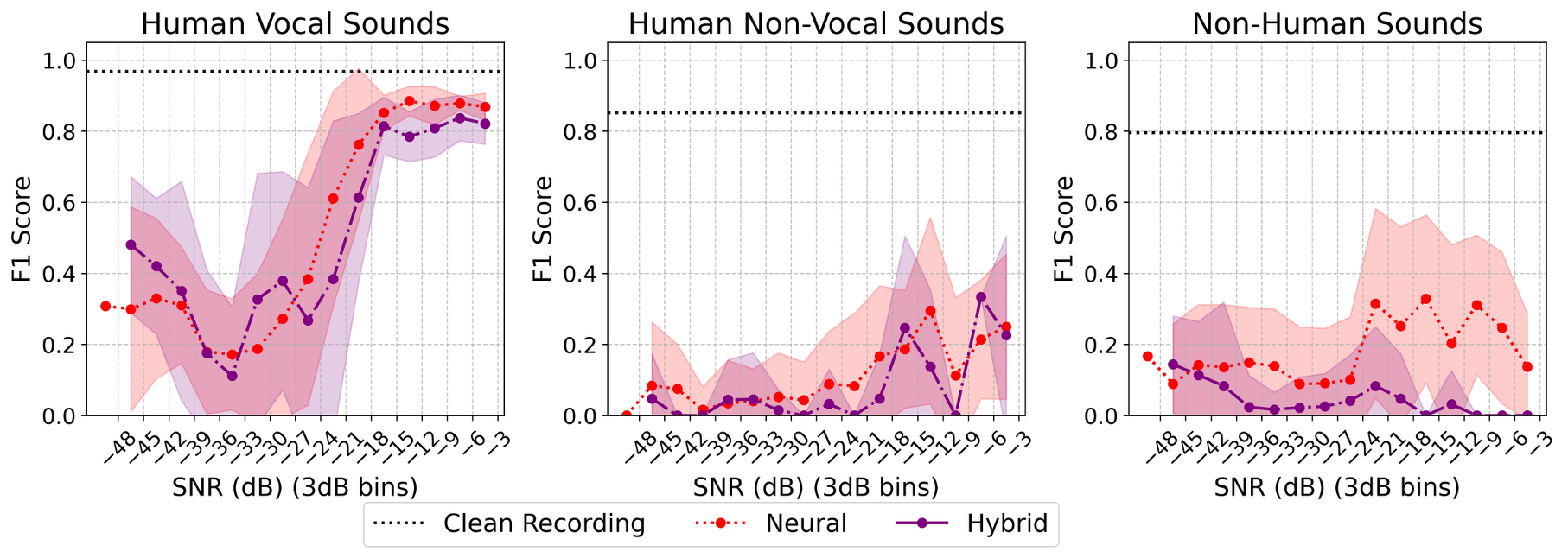}
    \caption{Classification performance of SSLAM on noise suppressed recordings \name at different SNRs.}
    \label{fig:app1classification}
    \end{minipage}
\end{figure}

Figure \ref{fig:app1noisesupp} compares the performance of noise suppression techniques, Traditional (Trad), Neural, and Hybrid, as well as unprocessed original audio, for different sources (human vocal, non-vocal, and non-human) in presence of drone noise across at every 3 dB change of SNR, measured using SI-SDR (dB). These plots provide a granular visualization of the trends in Table \ref{tab:combined_results}. Neural and Hybrid methods outperform Trad, particularly at lower SNRs. 

Figure \ref{fig:app1classification} evaluates the classification performance of SSLAM on the noise suppressed audio from the two best performing noise-suppression techniques Neural and Hybrid across at every 3 dB change of SNR. The plot focuses on HV, HNV, and NH sound classes, with F1-Score as the metric. The trends support the observations in Table \ref{tab:combined_results} that performance of classification of HV sounds is significantly better HNV and NH for both the classification techniques. The performance approaches clean recording performance for HV at high SNR values.

\subsection{Computational Resources for Evaluation}
All experiments were conducted on a Linux system (5.15.0-119-generic) with an x86\_64 architecture using Python 3.9.21. The hardware configuration consisted of a 36-thread CPU (18 physical cores @ 2.95 GHz), 134.7 GB of system RAM, and an NVIDIA GeForce RTX 3090 GPU with 24GB of VRAM (utilizing 58MB during idle measurements at 28°C). The computational resources required at each stage of the benchmarking pipeline is provided in Table \ref{tab:computationalresources}. Complete pipeline along with computational resources calculations is provided here: \url{https://github.com/augmented-human-lab/DroneAudioSet-code/blob/main/notebooks/overall_pipeline.ipynb}.
\begin{table}[]
    \centering
    \renewcommand{\arraystretch}{1}
    \resizebox{\textwidth}{!}{
    \begin{tabular}{|c|c|c|c|c|c|}
    \toprule
         \textbf{Module} &  \textbf{Exec.~time (sec)}&\textbf{Max CPU Usage (\%) }& \textbf{Memory Usage (MB)} & \textbf{Max GPU Usage (\%)} & \textbf{GPU Memory Usage (MB)}\\\midrule
         Beamforming (MVDR)&94.61&12.6&19900&0&- \\\hline
         Spectral Gating&5.52&2.6&728&0&-\\\hline
         MPSENET&31.35&3.5&704&100&13721\\\hline
         SSLAM&83.28&3.2&4216&19&819\\\hline
    \end{tabular}}
    \caption{Computational Resources needed by different modules to process six audio files of 12.6 min duration in total. Note, here the modules are sequentially applied, i.e. audio files are passed to beamforming, the output of which are passed through spectral gating, and so on.}
    \label{tab:computationalresources}
\end{table}
\section{\name Datasheet}
\label{app:datasheet}
Following the guidelines suggested by Gebru \textit{et al.}~\cite{gebru2021datasheets}, we document details of the \name dataset below.

\subsection{Motivation}

\textcolor{\sectioncolor}{\textbf{For what purpose was the dataset created?} Was there a specific task in mind? Was there a specific gap that needed to be filled? Please provide a description.}

We collect the dataset for human presence detection using audio in drone-based search and rescue. Existing drone audition datasets do not cover a diverse range of signal to noise ratios that represent realistic settings. Hence, we perform a more systematic data collection, covering a lot of scenarios including different source configurations (source sound types, loudness), microphone configurations (number, position) as well as drone configurations (type of drone, throttle level). 

\textcolor{\sectioncolor}{\textbf{Who created the dataset (e.g., which team, research group) and on behalf of which entity? (e.g., company, institution, organization)?}}

The dataset was collected by members of the Augmented Human Lab, at the National University of Singapore.

\textcolor{\sectioncolor}{\textbf{Who funded the creation of the dataset?}}

Ministry of Defence Singapore, MINDEF-DGA Joint Programme.

\textcolor{\sectioncolor}{\textbf{Any other comments?}}

None.

\subsection{Composition}

\textcolor{\sectioncolor}{\textbf{What do the instances that comprise the dataset represent? (e.g., documents, photos, people, countries)?} Are there multiple types of instances (e.g., movies, users, and ratings; people and interactions between them; nodes and edges)? Please provide a description.}

\name dataset includes multi-channel audio recordings, which can be split into three categories:
\begin{itemize}
    \item \textbf{Drone Noise + Sound Source Recordings.} This category denotes the largest segment of our dataset (15 hours out of 23.5 hours of recording duration). Here, we capture the drone noise as well as the source sounds simultaneously, while varying the drone, microphone and sound source configurations.
    \item \textbf{Drone-Only Recordings.} This category includes a total of 2.3 hours of data, which includes drone noise, with varying drone and microphone configurations. 
    \item \textbf{Source-Only Recordings.} This category consists of 6.2 hours of recordings, where we collect only the source sounds, while varying the source sound and microphone configurations. 
\end{itemize}

\textcolor{\sectioncolor}{\textbf{How many instances are there in total (of each type, if appropriate)?}}

Table~\ref{tbl:data-collection-summary} summarizes the recording duration for each configuration in our data collection. 

\textcolor{\sectioncolor}{\textbf{What data does each instance consist of?} “Raw” data (e.g., unprocessed text or images) or features? In either case, please provide a description.}

Our dataset consists of multi-channel audio recordings -- 8-channel audio recordings for microphones, \micup and \micdown, as well as single channel audio recordings corresponding to \miccenter, each sampled at 16~kHz.

\textcolor{\sectioncolor}{\textbf{Is there a label or target associated with each instance?} If so, please provide a description.}

All our recordings with source sounds have a mapping to indicate whether they are -- human vocal, human non-vocal, or non-human (ambient) sounds. In particular, we partition our ground-truth source audio into six files, notated as \texttt{File1} \dots \texttt{File6}, each containing a mix of different source types.  We provide annotations for all the six files at -- \url{https://huggingface.co/datasets/ahlab-drone-project/DroneAudioSet}, in the \texttt{ground-truth-annotations} folder.

\textcolor{\sectioncolor}{\textbf{Is any information missing from individual instances?} If so, please provide a description, explaining why this information is missing (e.g., because it was unavailable). This does not include intentionally removed information, but might include, e.g., redacted text.}

We do not have any missing information.

\textcolor{\sectioncolor}{\textbf{Are the relationships between individual instances made explicit (e.g., users’ movie ratings, social network links)?} If so, please describe how these relationships are made explicit.}

Yes, each audio recording file is annotated by a feature (called \texttt{path}) that exactly specifies all the configurations of the microphones, drones, and sound sources, thereby helping identifying different files with a certain configuration. 

\textcolor{\sectioncolor}{\textbf{Are there recommended datasplits (e.g., training, development/validation, testing)?} If so, please provide a description of these splits, explaining the rationale behind them.}

While in the current work we only perform inference on all the data, we provide recommendations for splits to facilitate future model development. Given that a good split should ensure sufficient diversity in the training, while testing on new data for testing, we aim to split our train/valid/tests based on the six files, six files, \texttt{File1} \dots \texttt{File6} which contain the three categories of source sounds (HV, NHV and NH), but different instances. In particular, all files except \texttt{File3} and \texttt{File4} contains 90~s, 30~s and 30~s, of each of the three categories, while \text{File3} only contains 90~s of HV, and \texttt{File4} only contains 30~s each of the other two. Overall, our recommendation is to consider the first four files (\texttt{File1} -- \texttt{File4}) for training, \texttt{File5} for validation, and \texttt{File6} for testing.

\textcolor{\sectioncolor}{\textbf{Are there any errors, sources of noise, or redundancies in the dataset?} If so, please provide a description.}

There are no errors to our knowledge.

\textcolor{\sectioncolor}{\textbf{Is the dataset self-contained, or does it link to or otherwise rely on external resources (e.g., websites, tweets, other datasets)?} If it links to or relies on external resources, a) are there guarantees that they will exist, and remain constant, overtime; b) are there official archival versions of the complete dataset (i.e., including the external resources as they existed at the time the dataset was created); c) are there any restrictions (e.g., licenses,fees) associated with any of the external resources that might apply to a dataset consumer? Please provide descriptions of all external resources and any restrictions associated with them, as well as links or other access points, as appropriate.}

Yes, the dataset is self-contained. 

\textcolor{\sectioncolor}{\textbf{Does the dataset contain data that might be considered confidential (e.g.,
data that is protected by legal privilege or by doctor-patient
confidentiality, data that includes the content of individuals’ non-public
communications)?
}
If so, please provide a description.}

No, the data collection process does not involve humans, hence there is no confidential data. All the source human sounds are gathered from various open-source data repositories, which are released under Creative Commons license.%

\textcolor{\sectioncolor}{\textbf{Does the dataset contain data that, if viewed directly, might be offensive,
insulting, threatening, or might otherwise cause anxiety?
} If so, please describe why.}

The dataset contains sounds of drones, human screams and baby cries which can induce distress and anxiety, if heard for long durations.

\textcolor{\sectioncolor}{\textbf{Does the dataset identify any subpopulations (e.g., by age, gender)?
} If so, please describe how these subpopulations are identified and provide a description of their respective distributions within the dataset.}

No, the data collection does not involve humans. 

\textcolor{\sectioncolor}{\textbf{Is it possible to identify individuals (i.e., one or more natural persons),
    either directly or indirectly (i.e., in combination with other data) from
    the dataset?}}

No.

\textcolor{\sectioncolor}{\textbf{Does the dataset contain data that might be considered sensitive in any way
(e.g., data that reveals racial or ethnic origins, sexual orientations,
religious beliefs, political opinions or union memberships, or locations;
financial or health data; biometric or genetic data; forms of government
identification, such as social security numbers; criminal history)?} If so, please provide a description.}

No. 

\textcolor{\sectioncolor}{\textbf{Any other comments?}}

None.

\subsection{Collection Process}

\textcolor{\sectioncolor}{\textbf{How was the data associated with each instance acquired?
} Was the data directly observable (e.g., raw text, movie ratings),
reported by subjects (e.g., survey responses), or indirectly
inferred/derived from other data (e.g., part-of-speech tags, model-based
guesses for age or language)? If data was reported by subjects or
indirectly inferred/derived from other data, was the data
validated/verified? If so, please describe how.}

\textcolor{\sectioncolor}{\textbf{What mechanisms or procedures were used to collect the data (e.g., hardware apparatus or sensor, manual human curation, software program, software API)?} How were these mechanisms or procedures validated?}

We explain the data collection process in detail in Section~\ref{sec:dataset}.

\textcolor{\sectioncolor}{\textbf{If the dataset is a sample from a larger set, what was the sampling
strategy (e.g., deterministic, probabilistic with specific sampling
probabilities)?
}}

Yes, the dataset is indeed sampled, given that there are endless options for drone types, drone throttles, microphone configurations and so on. We sample specific configurations that provide sufficient diversity. In particular, for drone types, we chose two drones with sufficiently different sizes and weights, to account for different acoustic properties. Similarly, for microphone configurations, we place two microphone arrays above and below the drone to capture sound from regions with varied turbulence due to propeller movement. We justify our choices for different parameters in Section~\ref{sec:dataset}.

\textcolor{\sectioncolor}{\textbf{Who was involved in the data collection process (e.g., students, crowdworkers, contractors) and how were they compensated (e.g., how much were crowdworkers paid)?
}}

Undergraduate students were hired under the university's student work scheme (NSWS) to assist with the data collection, and they were paid 17 SGD per hour, in accordance to the university's policies.

\textcolor{\sectioncolor}{\textbf{Over what timeframe was the data collected?
} Does this timeframe match the creation timeframe of the data associated
with the instances (e.g., recent crawl of old news articles)? If not,
please describe the timeframe in which the data associated with the
instances was created. Finally, list when the dataset was first published.}

The data was collected over a span of four months, from February 2025 to May 2025. The dataset was first published on 15 May 2025.  

\textcolor{\sectioncolor}{\textbf{Were any ethical review processes conducted (e.g., by an institutional
review board)?
}}

As there were no humans involved, there was no IRB involved. However, given that our experiments involved indoor drone experiments, we performed detailed risk assessment and got approval from the university's Office of Risk Management and Compliance.

\textcolor{\sectioncolor}{\textbf{Did you collect the data from the individuals in question directly, or obtain it via third parties or other sources (e.g., websites)?
}}

We only collected source sound files from public repositories, which do not have any identifiable information.

\textcolor{\sectioncolor}{\textbf{Were the individuals in question notified about the data collection?
}
If so, please describe (or show with screenshots or other information) how
notice was provided, and provide a link or other access point to, or
otherwise reproduce, the exact language of the notification itself.}

NA

\label{sec:consent}
\textcolor{\sectioncolor}
{\textbf{Did the individuals in question consent to the collection and use of their
data?} If so, please describe (or show with screenshots or other information) how
consent was requested and provided, and provide a link or other access
point to, or otherwise reproduce, the exact language to which the
individuals consented.}

NA

\textcolor{\sectioncolor}{\textbf{If consent was obtained, were the consenting individuals provided with a
mechanism to revoke their consent in the future or for certain uses?
} If so, please provide a description, as well as a link or other access point to the mechanism (if appropriate)}

NA

\textcolor{\sectioncolor}{\textbf{Has an analysis of the potential impact of the dataset and its use on data
subjects (e.g., a data protection impact analysis)been conducted?
}
If so, please provide a description of this analysis, including the
outcomes, as well as a link or other access point to any supporting
documentation.}

NA

\textcolor{\sectioncolor}{\textbf{Any other comments?}}

None.

\subsection{Preprocessing/Cleaning/Labeling}\label{app:datasheet_preprocessing}

\textcolor{\sectioncolor}{\textbf{Was any preprocessing/cleaning/labeling of the data
done (e.g.,discretization or bucketing, tokenization, part-of-speech
tagging, SIFT feature extraction, removal of instances, processing of
missing values)?
}
If so, please provide a description. If not, you may skip the remainder of
the questions in this section.}

We performed basic offset correction to remove zero-valued samples in the beginning of the audio recording, occasionally caused due to recording issues. We also resampled our collected audio data from 48~kHz to 16~kHz, as is preferred for most audio ML tasks. However, no additional pre-processing has been done beyond offset correction and resampling. 

\textcolor{\sectioncolor}{\textbf{Was the “raw” data saved in addition to the preprocessed/cleaned/labeled
data (e.g., to support unanticipated future uses)?}}

While we do not release the raw data, we save them locally in our servers. We are happy to share the raw data if there is such a requirement for a specific use case. Please contact the authors when you do so.

\textcolor{\sectioncolor}{\textbf{Is the software used to preprocess/clean/label the instances available?
}}

We share this code on our GitHub: \url{https://github.com/augmented-human-lab/DroneAudioSet-code/blob/main/scripts/0_preprocess_data.py}.

\textcolor{\sectioncolor}{\textbf{Any other comments?}}

None.

\subsection{Uses}

\textcolor{\sectioncolor}{\textbf{Has the dataset been used for any tasks already?
}
If so, please provide a description.}

We have only used the dataset for the two tasks mentioned in this paper -- detecting human-presence and providing recommendations for drone-audition. 

\textcolor{\sectioncolor}{\textbf{Is there a repository that links to any or all papers or systems that use the dataset?
}}

Once others begin to use this dataset and cite it, we will maintain a list at \url{https://apps.ahlab.org/DroneAudioSet-code/}.

\textcolor{\sectioncolor}{\textbf{What (other) tasks could the dataset be used for?
}}

We have listed out other application possibilities in Section~\ref{sec:discussion}.

\textcolor{\sectioncolor}{\textbf{Is there anything about the composition of the dataset or the way it was
collected and preprocessed/cleaned/labeled that might impact future uses?
}
For example, is there anything that a future user might need to know to
avoid uses that could result in unfair treatment of individuals or groups
(e.g., stereotyping, quality of service issues) or other undesirable harms
(e.g., financial harms, legal risks) If so, please provide a description.
Is there anything a future user could do to mitigate these undesirable
harms?}

The dataset has a limitation of being collected from a drone affixed to a frame, as opposed to real-world settings that involve hovering drones. Future users should ensure that they perform additional field tests with hovering drones.

\textcolor{\sectioncolor}{\textbf{Are there tasks for which the dataset should not be used?
}
If so, please provide a description.
}

The dataset should not be used for surveillance applications that lead to gathering private information (such as speech) without user consent. 

\textcolor{\sectioncolor}{\textbf{Any other comments?
}}

None

\subsection{Distribution}

\textcolor{\sectioncolor}{\textbf{Will the dataset be distributed to third parties outside of the entity
(e.g., company, institution, organization) on behalf of which the dataset
was created?
}
If so, please provide a description.
}

Yes, the dataset is publicly available and on the internet.

\textcolor{\sectioncolor}{\textbf{How will the dataset will be distributed (e.g., tarball on website, API,
GitHub)?
}
Does the dataset have a digital object identifier (DOI)?
}

The dataset is currently available at \url{https://huggingface.co/datasets/ahlab-drone-project/DroneAudioSet}.

\textcolor{\sectioncolor}{\textbf{When will the dataset be distributed?
}
}

The dataset is available for download at \url{https://huggingface.co/datasets/ahlab-drone-project/DroneAudioSet}. We will
await reviewer feedback before announcing this more publicly.

\textcolor{\sectioncolor}{\textbf{Will the dataset be distributed under a copyright or other intellectual
property (IP) license, and/or under applicable terms of use (ToU)?
}
If so, please describe this license and/or ToU, and provide a link or other
access point to, or otherwise reproduce, any relevant licensing terms or
ToU, as well as any fees associated with these restrictions.
}

We distribute it under the MIT License, as described at \url{https://opensource.org/license/mit/}.

\textcolor{\sectioncolor}{\textbf{Have any third parties imposed IP-based or other restrictions on the data
associated with the instances?
}
If so, please describe these restrictions, and provide a link or other
access point to, or otherwise reproduce, any relevant licensing terms, as
well as any fees associated with these restrictions.
}

We collect all data ourselves and therefore have no third parties with IP-based restrictions on the data.

\textcolor{\sectioncolor}{\textbf{Do any export controls or other regulatory restrictions apply to the
dataset or to individual instances?
}
If so, please describe these restrictions, and provide a link or other
access point to, or otherwise reproduce, any supporting documentation.
}

No.

\textcolor{\sectioncolor}{\textbf{Any other comments?
}}

None

\subsection{Maintenance}

\textcolor{\sectioncolor}{\textbf{Who is supporting/hosting/maintaining the dataset?
}
}

The dataset is hosted/maintained indefinitely in the Hugging Face Repository at \url{https://huggingface.co/datasets/ahlab-drone-project/DroneAudioSet}.

\textcolor{\sectioncolor}{\textbf{How can the owner/curator/manager of the dataset be contacted (e.g., email
address)?
}
}

Chitralekha Gupta can be contacted at \url{chitralekha@ahlab.org}, and Soundarya Ramesh can be contacted at \url{soundarya@ahlab.org}.

\textcolor{\sectioncolor}{\textbf{Is there an erratum?
}
}

There is currently no erratum, but if there are any errata in the future, we will publish them on the
website at \url{https://apps.ahlab.org/DroneAudioSet-code/}. 

\textcolor{\sectioncolor}{\textbf{Will the dataset be updated (e.g., to correct labeling errors, add new
instances, delete instances)?
}
If so, please describe how often, by whom, and how updates will be
communicated to users (e.g., mailing list, GitHub)?
}

To the extent that we notice errors, they will be fixed and the dataset will be updated.

\textcolor{\sectioncolor}{\textbf{If the dataset relates to people, are there applicable limits on the
retention of the data associated with the instances (e.g., were individuals
in question told that their data would be retained for a fixed period of
time and then deleted)?
}
If so, please describe these limits and explain how they will be enforced.
}

NA

\textcolor{\sectioncolor}{\textbf{Will older versions of the dataset continue to be
supported/hosted/maintained?
}
If so, please describe how. If not, please describe how its obsolescence
will be communicated to users.
}

We will maintain older versions of the dataset for consistency with the figures in this paper. These will be posted on a separate section on our website, in the event that older versions become necessary.

\textcolor{\sectioncolor}{\textbf{If others want to extend/augment/build on/contribute to the dataset, is
there a mechanism for them to do so?
}
If so, please provide a description. Will these contributions be
validated/verified? If so, please describe how. If not, why not? Is there a
process for communicating/distributing these contributions to other users?
If so, please provide a description.
}

We will accept extensions to the dataset as long as they follow the procedures we outline in this paper. We will label these extensions as such from the
rest of the dataset and credit those who collected the data. The authors of \name should be contacted about incorporating extensions.

\textcolor{\sectioncolor}{\textbf{Any other comments?
}}

None.

\end{document}